\documentclass[12pt]{article}
\usepackage[margin=0.95 in]{geometry}
\usepackage{amsmath} 
\usepackage{amssymb,amsfonts}
\usepackage[all]{xy}
\usepackage{graphicx}
\usepackage{simplewick}
\usepackage{amsmath}
\usepackage{amssymb}
\usepackage{float}
\usepackage{array}
\usepackage{tikz}
\usepackage{mathtools}
\usepackage{mathrsfs} 
\usepackage[hidelinks]{hyperref}
\usepackage{cite}
\numberwithin{equation}{section}
\usepackage[utf8x]{inputenc}

\newcommand{\be}{\begin{equation}}
\newcommand{\ee}{\end{equation}}
\def\bea{\begin{eqnarray}}
\def\eea{\end{eqnarray}}

\newcommand{\ii}{i}
\newcommand{\pa}{\partial}

\numberwithin{equation}{section}
\numberwithin{table}{section}

\begin{document}

\hypersetup{pageanchor=false}
\begin{titlepage}
\vbox{
    \halign{#\hfil         \cr
           } 
      }  
\vspace*{15mm}
\begin{center}
{\Large \bf 
The $AdS_3 \times S^1$ Chiral Ring
}

\vspace*{10mm}

{\large Sujay K. Ashok$^a$, 
Songyuan Li$^b$, and Jan Troost$^b$}
\vspace*{8mm}

$^a$The Institute of Mathematical Sciences, \\
          Homi Bhabha National Institute (HBNI),\\
		 IV Cross Road, C.I.T. Campus, \\
	 Taramani, Chennai, India 600113

\vspace{.8cm}

$^b$ Laboratoire de Physique de l'\'Ecole Normale Sup\'erieure \\ 
 \hskip -.05cm
 CNRS, ENS, Universit\'e PSL,  Sorbonne Universit\'e, \\
 Universit\'e de Paris 
 \hskip -.05cm F-75005 Paris, France	 
\vspace*{0.8cm}
\end{center}

\begin{abstract}
We study $AdS_3 \times S^1 \times Y$ supersymmetric string theory backgrounds with Neveu-Schwarz-Neveu-Schwarz flux that are dual to ${\cal N}=2$ superconformal theories on the boundary. We classify all worldsheet vertex operators that correspond to space-time chiral primaries. We compute space-time chiral ring structure constants for operators in the zero spectral flow sector using the operator product expansion in the worldsheet theory. We find that
the structure constants take a universal form that depends only on the topological data of the ${\cal N}=2$ superconformal theory on $Y$.

\end{abstract}

\end{titlepage}
\hypersetup{pageanchor=true}

\setcounter{tocdepth}{2}
\tableofcontents

\section{Introduction}
Holographic duality is likely a generic property of quantum gravity \cite{tHooft:1993dmi,Susskind:1994vu}. While convincing examples of the duality are known \cite{Maldacena:1997re}, the duality remains ill-understood. Understanding holographic duality in detail could imply resolving the information paradox, or finding a unitary quantization of gravity in de Sitter space-time. Thus, there is promise in understanding holographic dualities in greater depth. For instance, one can enlarge the class of examples of holographic dualities, or attempt to understand those examples that we know in more detail.

The holographic string theory backgrounds that are most easily exactly quantized in the inverse curvature expansion in the bulk may be the three-dimensional anti-de Sitter backgrounds with Neveu-Schwarz-Neveu-Schwarz flux. The world sheet model is 
a Wess-Zumino-Witten model on the universal cover of the non-compact group $SL(2,\mathbb{R})$, and as such may be exactly solvable. Indeed, a lot of progress has been made in explicitly solving the theory (see e.g.  \cite{Kutasov:1999xu,Maldacena:2001km,Dei:2021xgh}). Moreover, in the presence of extended supersymmetry, the boundary conformal field theory can be  twisted to give rise to a topological conformal field theory. It is a reasonable goal to attempt to prove the holographic duality between the topologically twisted bulk theory and the topological conformal field theory on the boundary in this setting \cite{Li:2020nei}. 

String theories in three-dimensional anti-de Sitter space with NS-NS flux admit ${\cal N}=2$ superconformal symmetry when they are of the form $AdS_3 \times S^1 \times Y$ where $Y$ is a  $N=2$ world sheet superconformal field theory \cite{Giveon:2003ku}. We determine the full spectrum of spacetime chiral primaries in these backgrounds in section \ref{SpaceTimeChiralPrimaries}. We compute structure constants of the space-time chiral ring in section \ref{SpaceTimeChiralRing}. Our technique is to compute the space-time operator products directly, using world sheet techniques. The underlying motivation is to perform the calculation in the bulk in a spirit that will eventually allow for topological twisting. The reward we reap  is that the calculation becomes simple and since it is only based on the necessary properties of the ${\cal N}=2$ superconformal background, it is generally valid. We thus  extend calculations performed in backgrounds with ${\cal N}=4$ superconformal symmetry \cite{Dabholkar:2007ey,Gaberdiel:2007vu} that provided powerful checks of the holographic correspondence.  The ${\cal N}=2$ superconformal backgrounds are less constrained and could exhibit a richer set of structure constants. 
We discuss our results and conclude in section \ref{Conclusions}. In a set of appendices we summarize many details regarding the worldsheet conformal field theory on the $AdS_3\times S^1\times Y$ background, and in particular, lay bare some generic properties of $N=2$ superconformal world sheet theories.

\section{The Space-time Chiral Primaries}
\label{SpaceTimeChiralPrimaries}
In this section, we classify the spectrum of space-time chiral primary fields in $AdS_3$ backgrounds with ${\cal N}=2$ spacetime superconformal symmetry. The classification extends results obtained in various approaches in references \cite{Giveon:1999zm,Argurio:2000tb,Eberhardt:2017fsi,Ashok:2020dnc} among others. The chiral primaries form a basis for the chiral ring \cite{Lerche:1989uy}, a topological and non-trivial observable of the extended superconformal field theory. 
\subsection{The Extended Superconformal Backgrounds}
We study a class of string theory backgrounds with spacetime ${\cal N}=2$ superconformal symmetry. When these backgrounds carry only NS-NS flux and can be described by a world sheet conformal field theory, they were argued to take the form $AdS_3 \times S^1 \times Y$ \cite{Giveon:2003ku} where the factor $Y$ corresponds to a world sheet  theory with $N=2$ world sheet conformal symmetry.\footnote{It is clear that one can also allow orbifolds thereof and that there may be further wiggle room (e.g. by demanding only chiral ${\cal N}=2$ symmetry). Still,  the  $AdS_3 \times S^1$ geometry is a natural geometric realization of extended superconformal symmetry.} The extended spacetime supersymmetry forces the presence of a $U(1)_R$ symmetry that in turn is geometrically incarnated as a circle factor in the spacetime. These backgrounds provide a large class of theories in which to study holography with extended supersymmetry and thus provide a fertile testing ground for attempting to prove a topologically twisted version of the correspondence between quantum gravity in anti-de Sitter space and conformal field theory \cite{Li:2020nei}. 

The world sheet central charge of the $AdS_3 \times S^1 \times Y$ theory is critical when it satisfies
\begin{equation}
c_{tot} = 3+\frac6k + 1 + 4 \times \frac12 + c_Y = 15~.
\end{equation}
We have included four world sheet fermion partners for the $AdS_3 \times S^1$ bosons. 
The level $k$ is equal to the square of the $AdS_3$ cosmological constant length scale divided by $\alpha'$ (which in turn equals $2 \pi$ times the fundamental string tension). We conclude that the critical world sheet model contains a factor theory $Y$ with central charge 
\be
c_Y = 9 - \frac{6}{k} \, .
\ee 
We take the radius of the circle  equal to $R=\sqrt{k \alpha'}$ which equals the radius of curvature of the $AdS_3$ space-time.

\subsection{The  Spectrum }
We work in the Neveu-Schwarz-Ramond  formalism on the world sheet and in a first instance ignore light-cone oscillator degrees of freedom. We consider vertex operators that factorize according to the spacetime factors.  The world sheet conformal dimension $L_0$, the circle left momentum $p_L$, the space-time energy $H$ and the space-time R-charge $Q_{\text{st}}$ are given by the   expressions -- see e.g. \cite{Ashok:2020dnc} for the detailed background --:
\begin{align}
\label{allquantumnos}
L_0&=  -\frac{h(h-1)}{k}- w (h+r) - \frac{(k+2)w^2}{4} +\frac12(f+a)^2 + \frac{p_L^2}{2}  +h_{Y}+ N~,
\\
p_L &= \frac{n'}{R}+ \frac{w' R}{\alpha'} = \frac{n'}{\sqrt{2k}} + \sqrt{\frac{k}{2}} w'
= \frac{1}{\sqrt{2k}} (n' + k w')
\\
H &= h+r+\frac{k+2}{2} w+f+a
\\
Q_{\text{st}} &= \sqrt{2k} p_L
\end{align}
The quantum numbers appearing in these equations are as follows.  First of all, we put $a=0$ for the world sheet NS sector and $a=-\frac12$ in the Ramond sector. The quantum numbers $(h,r,w)$ arise from the $sl(2,\mathbb{R})$ world sheet theory and $h$ is the world sheet $sl(2,\mathbb{R})$ spin, $w$ parameterizes the spectral flow and $r$ measures an elliptic $sl(2,\mathbb{R})$ spin component. We restrict to discrete representations of $sl(2,\mathbb{R})$ and will set $r=0$  -- see  \cite{Ashok:2020dnc} for a detailed justification. The super ghosts cancel  one of the two pairs of fermions and the quantum number $f$ corresponds to the fermion number arising from the remaining pair of fermions in the $AdS_3\times S^1$ sector -- one takes them to lie in the spatial $AdS_3$ directions. The quantum numbers $(n',w')$ are the momentum and winding along the circle direction. The conformal dimension of the operator arising from the compact conformal field theory $Y$ is denoted $h_Y$ and $N$ is the oscillator contribution to the conformal dimension. The momentum $p_L$ is determined in terms of the momentum and winding along the circle $S^1$.  
The spacetime energy $H$ is the $J_0^3$ time translation eigenvalue in the (supersymmetric) $AdS_3$ theory while the spacetime $R$-charge is proportional to the left-moving momentum along the $S^1$. Similar formulas hold in the right-moving sector. The space-time spectrum is determined by solving on-shell constraints (among others on $L_0$) in the Hilbert space of the world sheet conformal field theories.

\subsection{The Spacetime Chiral Primaries}

Our goal is to classify the single fundamental string states that are spacetime chiral primaries. These states satisfy the equality \cite{Lerche:1989uy}:
\be
2H = Q_{\text{st}}~. \label{ChiralPrimary}
\ee
As physical states, they are subject to the on-shell condition $L_0=\frac12$. In a first step, we  show that the space-time chiral primary equation (\ref{ChiralPrimary}) and the world sheet on-shell condition can be rendered independent of the spectral flow parameter $w$  by shifting the world sheet fermion number $f=f'-w$, or in other words, by applying spectral flow to the fermion sector as well. Firstly, the chiral primary condition (\ref{ChiralPrimary}) allows us to write the left circle momentum $p_L$ entirely in terms of the $AdS_3$ quantum numbers. Substituting this into the equation for the world sheet conformal dimension we obtain
\be
\label{L0forCP}
L_0 = \frac{h(1+2a+2f')}{k}+\frac{(k+2)}{2k}(f'+a)^2+ h_Y + N \, .
\ee
We see that indeed, the condition is independent of the spectral flow parameter $w$.

Spacetime chiral primaries can arise from  both the world sheet  Ramond and NS sectors and we discuss each of these in turn. For simplicity, we work chirally -- the physical spectrum is obtained by suitably combining left and right moving world sheet excitations in a modular invariant, GSO projected theory. 

\subsubsection{The Ramond Sector}
 In the Ramond sector we set the parameter $a=-1/2$. Substituting this into the world sheet conformal dimension $L_0$ \eqref{L0forCP} we find that it is minimized for fermion number $f'=0$. We attain the minimum: 
\begin{align}
L_0 
&=\frac{k+2}{8k} +h_{Y} +N~.
\end{align}
In the Ramond sector of the theory $Y$, the  conformal dimension is bounded from below:
\be
h_{Y} \ge \frac{c_{Y}}{24}=\frac{3}{8} - \frac{1}{4k}~.
\ee
For a space-time chiral primary state, we therefore find that the world sheet conformal dimension satisfies $L_0 \ge 1/2$. The physical state condition $L_0=1/2$ is satisfied when there are no oscillator excitations and {\it  only} for Ramond ground states of the theory $Y$.

There are two more constraints to be satisfied  involving  the spacetime and world sheet R-charges. The first is that the spacetime R-charge is given by the momentum and winding on the circle. Combining this with the spacetime chiral primary condition we find:
\begin{align}
2H = 2h-1+kw &= Q_{\text{st}} = n'+kw' \, , \label{SpinQuantization}
\end{align}
where we recall that $w$ is the spectral flow number and $(n',w')$ are the momentum and winding on the circle. 
The constraint can be rewritten in the form
\be
2h-1 = n'+k(w'-w)~,
\ee
which states that the combination of spin $2h-1$ is an integer plus an integer multiple of $k$. For discrete states in $AdS_3$ the spins are constrained to lie in the interval $2h-1 \in (0, k]$. We conclude that there are only a finite number of spins $h$ allowed.
The last constraint in the classification problem of space-time chiral primaries in the world sheet Ramond sector is to implement the world sheet GSO projection. We discuss this constraint in subsection \ref{CovariantRamond}.  
\subsubsection{The Neveu-Schwarz Sector}

In the NS sector we set $a=0$ and the world sheet conformal dimension takes the form
\be
L_0= \frac{(2f'+1)h}{k}+ \frac{k+2}{2k} (f')^2 + h_Y +N~.
\ee
Extremizing with respect to the fermion number $f'$ (while ignoring the mild implicit GSO dependence of the dimension $h_Y$ on $f'$), we find that the fermion number $f'$ takes the values $0$ or $-1$. 
\paragraph{Case 1:} For the case
$f'=0$ we obtain the world sheet operator dimension 
$L_0=\frac{h}{k}+h_Y$. The on-shell value is still $L_0=1/2$.  Using the constraint on the allowed range of the spin $h$ we obtain a bound on the conformal dimension $h_Y$ of the operator factor in the theory $Y$:\footnote{The bound on the spin $h$ gives a weaker lower bound. Unitarity of the $Y$-theory implies that $h_Y\ge 0$.}  
\be 
\label{hyrange1}
0 \le h_Y \le \frac12 -\frac{1}{2k}~.
\ee
Furthermore, we  need to implement the 
GSO constraint:
\begin{equation}
q_{\text{ws}} = 1- 2 h_Y +q_Y \in \text{odd integers} \, .
\end{equation}
By combining the GSO constraint with the bound on the dimension $h_Y$ we find that for a fixed world sheet R-charge $q_Y$ only one dimension $h_Y$ can solve the GSO constraint (because the range of $2h_Y$ is smaller than one). The unitarity of the $N=2$ superconformal theory on $Y$ implies the inequality $2h_Y \ge |q_Y|$. For positive world sheet R-charge, $q_Y \ge 0$, there are only the primaries satisfying $2h_Y=q_Y+2 \mathbb{Z}$ that can therefore solve the GSO constraint. Moreover, if the charge $q_Y$ is positive, then by the constraint on $2h_Y$, which is less than one, we must have $2h_Y=q_Y$. We conclude that world sheet chiral primaries are the only states that satisfy the constraint in case the world sheet R-charge in the theory $Y$ is positive.

For negative R-charge, we have $2h_Y \ge -q_Y$, or $-2h_Y \le q_Y$.  We conclude that:
$-q_Y \le 1-1/k$ in order for $2 h_Y$ to exist. We therefore have the constraint
\begin{equation}
0 \ge q_Y \ge -1+1/k \, .
\end{equation}
We add the bound on $1-2h_Y$ and on $q_Y$ to find 
that: 
\begin{equation}
-1+2/k \le 1-2h_Y+q_Y < 1+1/k \, .
\end{equation}
For this to be an odd integer, we need $2h_Y=q_Y$ which implies that both are zero in a unitary theory $Y$. Thus, this is the identity operator in the unitary conformal field theory $Y$. It also falls into the previous category, with positive R-charge. Therefore, there are no extra space-time chiral primaries in this category. 
The only solutions to all the constraints that can give rise to a spacetime chiral primary is a world sheet chiral primary with conformal dimension $h_Y$ in the range  \eqref{hyrange1}.

\paragraph{Case 2:} 
For the case $f'=-1$ we find $L_0 =\frac12+\frac{1-h}{k} +h_Y=\frac12$ for the on-shell constraint.
Solving this constraint and using the range of spins $h$, we find the same range \eqref{hyrange1} for the conformal dimension:
\begin{align}
0 \le h_Y \le \frac12 -\frac{1}{2k}
\end{align}
The GSO constraint in this case is given by 
\begin{equation}
q_{\text{ws}} = -1+2h_Y+q_Y 
\in \text{odd integers} \, .
\end{equation}
We  proceed as before and observe that for a given charge $q_Y$ there is only a single value of $h_Y$ that can solve the GSO constraint. Using the unitarity bound of the theory $Y$, let us discuss the two cases of positive and negative R-charge. For $q_Y\le 0$ only primaries satisfying $2h_Y = -q_Y + 2\mathbb{Z}$ can solve the GSO constraint. Given the constraint on $2h_Y$, we must therefore have a primary with $2h_Y = - q_Y$. Thus only world sheet anti-chiral primaries satisfy all the constraints in this case. 
For $q_Y \ge 0$ we have the unitarity constraint $2h_Y \ge q_Y$. The bound on $h_Y$ thus implies a bound on $q_Y$:
\be
0 \le q_Y \le 1-\frac1k~.
\ee
Adding this to the bound on $2h_Y$ we obtain 
\be
-1\le  -1 + 2h_Y  + q_Y \le 1-\frac2k~.
\ee
This is the same combination that occurs in the expression of the world sheet R-charge and for this to be an odd integer we must have $2h_Y + q_Y = 0$. In conjunction with the unitarity bound for $q_Y \ge 0$, the only possibility is that both the dimension and the R-charge vanish.  Thus, in the second case, only the anti-chiral primaries in the world sheet theory $Y$ give rise to space-time chiral primaries.

Finally, let us point out that there is a strong upper bound on the conformal dimension of an operator in the theory $Y$ that can feature in a space-time chiral primary. The upper bound on its conformal dimension is 
\be h_Y \le \frac12 - \frac{1}{2k} \, ,
\ee
and the upper bound on the absolute value of its world sheet R-charge is therefore $|q_Y| \le 1-1/k$.
Note that it is smaller than one. Yet, the central charge of the theory $Y$ is $c_Y=9-6/k$ and therefore the bound is considerably stronger than the unitarity bound within the world sheet theory $Y$. 

\subsubsection*{A Corollary}
We observe a distinction between the range in which the cosmological length scale is larger than the string scale, $k \ge 1$, and in which it is smaller, $k <1$. In the latter case, we cannot obtain spacetime chiral primaries from the world sheet NS sector for a unitary theory $Y$ because of the bound  (\ref{hyrange1}).

\subsection{The Covariant Chiral Primary Vertex Operators}

So far we worked out the spectrum of spacetime chiral primaries in the absence of light cone oscillator excitations. That method provides a rather direct access to the spectrum of physical excitations.  For the purpose of computing the spacetime operator product expansions, it is however useful to write the vertex operators for these states in a covariant form. 

\subsubsection{The Neveu-Schwarz Sector}
The covariant vertex operators in the left moving Neveu-Schwarz sector in picture number $(-1)$ take the form \cite{Friedan:1985ge,Polchinski:1998rr}:
\begin{equation}
O^{(-1)} = c ~ e^{-\phi}~ O_{\text{matter}} \, .
\end{equation}
The world sheet conformal dimension has the same expression as before, but  we allow for light cone bosonic and fermionic oscillator excitations. The conformal dimension of the $c$ ghost is $-1$, and the $e^{-\phi}$ ghost excitation has dimension $1/2$. Therefore the matter operator of the unintegrated vertex operator remains of  dimension $1/2$. Light cone oscillator excitations in the matter operator would only raise the world sheet conformal dimension and are therefore incompatible with the space-time chiral primary constraint. 
We still need to check whether our excitation is BRST closed, and whether there are excitations that are BRST trivial. Given the light-cone analysis of spacetime chiral primaries, we suspect that the BRST cohomology comprises of all the states we found previously and none other. 

In order to check this, we need to compute the action of $Q_{BRST} = 1/(2 \pi i) \oint (cT+\gamma G)$  on these states. The first term gives zero because of the on-shell constraint $L_0=\frac12$ (and the absence of oscillators). The second term will only act when we have a non-trivial fermion excitation, and from our results, this can only be the fermion $\psi^-$ in the transverse $AdS_3$ directions that lowers the space-time conformal dimension. If such a fermion is present, the condition for BRST closedness will be that the $sl(2,\mathbb{R})$ lowering operator $j^-_0$ is zero on the bosonic state, i.e. that it is a $j^3_0$ lowest weight state. Our choice $r=0$ for the $j^3_0$ momentum of the discrete $sl(2,\mathbb{R})$ representations guarantees that this is indeed the case -- it precisely cements the choice of a lowest weight state in a discrete $D^+$ representation \cite{Ashok:2020dnc}. Moreover, in the compact theory $Y$, (anti-)chiral primaries are automatically annihilated by $G_{+1/2}$ and therefore, indeed, all the covariant states that we identified are BRST closed. 

To prove that none of these states is BRST exact, it is sufficient to recall that the BRST charge does not change the space-time conformal dimension or charge, and therefore any relevant seed would need to be a space-time chiral primary. We have just proven that those are all BRST closed and therefore they cannot give rise to  non-trivial BRST exact states. 

We conclude that we have two classes of covariant NS sector states. The first class has $H=h + \frac{kw}{2}$ and in the left-moving sector it is given by the  vertex operator: 
\begin{equation}
\label{cpns1defn}
O^{a\, (-1)}_{\text{NS}_1} = c ~  e^{-\phi}~ \Phi_{h}^w ~
\psi_w
~e^{\ii \sqrt{\frac{2}{k}}(h + \frac{kw}{2}) X}
~V^a_{c.p.}~.
\end{equation}
The left-moving circle momentum is $p_L =\sqrt{\frac{2}{k}}(h + \frac{kw}{2}) $ and the conformal dimension of the chiral primary in the $Y$ sector is determined from the on-shell condition to be
\be 
h_Y^a =\frac12 - \frac{h}{k}~. 
\ee
We denote the chiral ring elements by $V^a_{c.p.}$ where $a=1, \ldots \text{dim}({\cal R}_Y)$, where ${\cal R}_Y$ is the chiral ring of the ${\cal N}=2$ superconformal world sheet conformal field theory  $Y$.

Finally, we make a few remarks about the spectrally flowed operators we have defined. The spectral flow in the bosonic $sl(2,\mathbb{R})$ sector is analyzed in \cite{Maldacena:2000hw}. The spacetime conformal dimension associated to the flowed states is  listed in equation \eqref{allquantumnos}. The spectral flow in the fermionic sector is described in detail in \cite{Giribet:2007wp}. We have denoted the vacuum in this spectrally flowed fermionic sector $|\tilde 0\rangle$ as being created by the operator $\psi_w$. It  corresponds to the state
\be
\label{fermiongsw}
\psi_w \longleftrightarrow |\tilde 0\rangle = \frac{1}{k^{w/2}} \psi^-_{-w+\frac12}\psi^-_{-w+\frac32} \cdots \psi^-_{-\frac12}|0\rangle~.
\ee
It has world sheet fermion number $-w$ and world sheet conformal dimension $\frac{w^2}{2}$. It has zero spacetime R-charge and spacetime conformal dimension equal to $-w$. 

The second class of NS sector states consists of states with space-time conformal dimension $H= h + \frac{kw}{2} -1$ and the left-moving vertex operator takes the  form
\begin{equation}
O^{\bar a\, (-1)}_{\text{NS}_2}  =c~ e^{-\phi}~ \Phi_{h}^{w} ~\psi_{w+1}
~e^{\ii \sqrt{\frac{2}{k}}(h+\frac{kw}{2} - 1) X} 
~V^{\bar a}_{a.c.p.}~. 
\end{equation}
The momentum along the circle is determined in terms of the spacetime R-charge: $p_L=\sqrt{\frac{2}{k}}(h+\frac{kw}{2} - 1)$. 
The world sheet conformal dimension of the anti-chiral primary state in the sector $Y$ is
\be
 h_Y^{\bar a} = \frac{h-1}{k} ~.
\ee
We have denoted the anti-chiral primaries by $V^{\bar a}_{a.c.p.}$, where the index $\bar a$ again labels world sheet ring elements. We will assume that we have a left world sheet $U(1)_R$ conjugation symmetry and that the left chiral ring is isomorphic to the left anti-chiral ring.

We have denoted by $\psi_{w+1}$ the single fermionic excitation on top of the ground state  \eqref{fermiongsw}: 
\be
\label{psiw+1defn}
\psi_{w+1} \longleftrightarrow \frac{1}{\sqrt{k}} \psi^{-}_{-w-\frac12}|\tilde 0\rangle = \frac{1}{k^{(w+1)/2}} \psi^{-}_{-w-\frac12}\psi^-_{-w+\frac12} \cdots \psi^-_{-\frac12}|0\rangle~.
\ee
The world sheet and spacetime quantum numbers shift appropriately by the addition of this extra fermion.

\subsubsection{The Ramond Sector}
\label{covariantR}
\label{CovariantRamond}
We  perform a covariant analysis in the left-moving Ramond sector as well -- it is to be combined with a right-moving Ramond or Neveu-Schwarz sector. We consider the covariant vertex operators in the $(-\frac12)$ picture:  
\begin{equation}
O_{\text R}^{(-\frac12)} = c~ 
e^{-\frac{\phi}{2}}~
O_{\text{matter}} \,
\end{equation}
where $O_{\text{matter}}$ is of conformal dimension $5/8$.
The matter vertex operators that we identified previously are again the pertinent ones, but we now take into account the contribution of the fermionic zero modes  in the light-cone directions to the world sheet conformal dimension. We can for instance set the light cone fermion number to $s_0=+1/2$ (as in flat space superstring theory \cite{Polchinski:1998rr}). Effectively, we are then looking for solutions of $L_0=5/8-1/8=1/2$, and our problem reduces to the one that we solved previously. 

Again, we need to check how the BRST cohomology interfaces with our solution set. The $L_0$ constraint is once more taken care of while the fermionic part of the BRST operator also annihilates the state  because of the fact that we have solved the Dirac equation by picking signs of light cone momenta as well as $s_0=1/2$ -- see \cite{Polchinski:1998rr} as well as below. There are  no BRST exact states by the reasoning we described above. 

We summarize our results for the covariant Ramond sector vertex operator in picture $(-\frac12)$. 
The left-moving part of the matter vertex operator  has conformal dimension $H=h+\frac{kw}{2}-\frac12$ and takes the factorized form:
\begin{equation}
\label{vertopR}
O_{\text{matter}} = \Phi_{h}^{w} 
~e^{\ii \sqrt{\frac{2}{k}}(h+\frac{kw}{2}-\frac12) X}
~P_{\text{matter}} \, .
\end{equation}
The momentum along the $S^1$ direction is given by $p_L = \sqrt{\frac{2}{k}}(h+\frac{kw}{2}-\frac12)$. 
Let us be a bit more precise about the spin and internal fields that make up the operator $P_{\text{matter}}$. We refer to the discussion in \cite{Giveon:2003ku} and  Appendix \ref{wstheory} for further details. The four fermions that are the world sheet superpartners of the $AdS_3\times S^1$ directions are paired up and bosonized in order to write the vertex operators in the Ramond sector. We have 
\be
\label{Hcurrents}
\pa \hat H_1 =  \frac{2}{k}\psi^1\psi^2~, \qquad \pa \hat H_0 = -\ii \sqrt{\frac{2}{k}} \psi^0\psi^3~.
\ee
The hatted bosons have been defined in equation \eqref{hattedbosons} -- appropriate  cocycle factors have been included. The remaining factor in the matter operator can then be written as:
\be
P_{\text{matter}} =  e^{-\ii(\frac12 +w)\hat H_1 } e^{\pm \frac{\ii}{2} \hat H_0}~ \Sigma^{\bar a} ~,
\ee
where $\Sigma^{\bar a}$ are the Ramond ground state operators  of the ${\mathcal N}=2$  superconformal theory  $Y$. The label $\bar a$ runs over the Ramond ground states of the theory.  As explained in \cite{Lerche:1989uy, Cecotti:1991me} it is possible to realize the Ramond ground states either by acting with the chiral ring elements on the Ramond ground state with the lowest R-charge or equivalently by acting with the anti-chiral ring elements on the Ramond ground state with the highest R-charge. Thus, the ground states can be labelled by either $a$ or $\bar a$. Our choice of labelling the states by the anti-chiral primaries anticipates calculations to come.

The on-shell condition requires us to impose that the world sheet $N=1$ super current mode $G_0$ acting on the state is zero. The super current splits into three terms corresponding to the factored $AdS_3 \times S^1 \times Y$ model:
\be
G = \frac{2}{k}\psi^A J_A +\ii\,  \psi^{0} \partial X +G^Y ~.  \label{ThreeTermG}
\ee
When we act with the supercurrent on the vertex operator  \eqref{vertopR}, the coefficient of the simple pole is proportional to:
\begin{align}
p_L \gamma^0 + \frac{2}{k} \gamma^3( h+ \frac{kw}{2}-\frac12) 
&= \sqrt{\frac{2}{k}} (h+\frac{kw}{2} - \frac12)\gamma^0(1+\sqrt{\frac{2}{k}} \gamma^0\gamma^3)
\end{align}
where the $\gamma^\mu $ matrices represent the fermion zero modes.
The right hand side vanishes only if the coefficient of the field $\hat H_0$ in the exponent of the vertex operator is  $\frac12$, as we assumed previously. 
Finally we  impose a GSO projection on the vertex operator.  We write the Ramond sector ground state as (see Appendix \ref{YTheory} for details):
\be
\Sigma^{\bar a} = e^{\ii \sqrt{\frac{3}{c_Y}} q^{\bar a}_R Z}\, \Pi^{\bar a}~,
\ee
where $q^{\bar a}_R$ is the R-charge of the Ramond ground state in the Y-theory. Here $Z$ bosonizes the U$(1)_R$ charge of the $Y$ theory and we have separated out the factor that carries the world sheet R-charge. The chiral GSO projection is equivalent to the condition of locality with respect to the spacetime  supersymmetry generators. Following \cite{Giveon:2003ku} we choose a set of space-time supersymmetry generators (see equation \eqref{spacetimeQvop} in Appendix \ref{SpacetimeSusyAlgebra}), whose world sheet operators are given by:
\be
e^{-\frac{\phi}{2}} S_r^{\pm} = e^{-\frac{\phi}{2}}~ e^{-\ii r(\hat H_1\mp \hat H_0)}~ e^{\pm \ii \frac{X}{\sqrt{2k}}}~ e^{\pm \ii \sqrt{\frac{c_Y}{12}} Z} ~,
\ee
where $r= \pm\frac12$. We use the standard free field operator product expansions 
\begin{align}
e^{-\frac{\phi}{2}}\cdot e^{-\frac{\phi}{2}} &\approx \frac{e^{-\phi}}{z^{\frac14}}\cr
e^{-\ii r \hat H_1} \cdot e^{-\ii(\frac12+w)\hat H_1 }&\approx {z^{r(w+\frac12)}} {e^{-\ii(r+\frac12+w)\hat H_1 }}\cr
e^{\pm \ii r \hat H_0} \cdot e^{\frac{\ii}{2} \hat H_0} &\approx  {z^{\pm \frac{r}{2}}} {e^{\ii(\pm r+\frac12)\hat H_0}}
\\
e^{\pm \ii \frac{X}{\sqrt{2k}}}\cdot 
e^{ \frac{\ii}{\sqrt{2k}}(2h-1+kw) X}
&\approx {z^{\pm \frac{2h-1+kw}{2k} }} {e^{ \frac{\ii}{\sqrt{2k}}(2h-1+kw\pm 1) X}}\cr
e^{\pm \ii \sqrt{\frac{c_Y}{12}} Z}\cdot e^{\ii \sqrt{\frac{3}{c_Y}} q^{\bar a}_R Z}&\approx 
{z^{\pm \frac{q^{\bar a}_R}{2}}}
{e^{\ii \sqrt{\frac{3}{c_Y}} (q^{\bar a}_R\pm \frac{c_Y}{6}) Z}}~,
\nonumber
\end{align}
to show that  the locality of the operator product expansion amounts to the  constraint:
\be
-\frac12 +r(1\pm 1) + 2w(r\pm \frac12)\pm (\frac{2h-1}{k} + q^{\bar a}_R)\in~ \text{odd integers}.
\ee
Since $r\pm \frac12$ is always an integer,  multiplying by $2w$ always equals an even integer. Thus the $w$-dependent term  drops out from the oddness constraint. For both choices of signs the dependence on the index $r$ on the space-time supercharges  trivializes as well and we must satisfy:
\be
\label{wsGSOtwo}
\frac12 + \frac{2h-1}{k} + q^{\bar a}_R
\in \text{odd integers}
\ee
Thus locality with all four supercharges is ensured by satisfying the charge constraint  \eqref{wsGSOtwo}.

The classification problem has  been reduced to one that depends solely on the level $k$ and the nature of the world sheet conformal field theory $Y$. The number of solutions we find will depend on the spectrum of world sheet R-charges $q_Y$. A general constraint on that spectrum is that it is bounded by the central charge of the theory $Y$ \cite{Lerche:1989uy}:
\be
-\frac32+\frac1k \le q_R^{\bar{a}} \le \frac32-\frac1k~.
\label{GeneralBound}
\ee
Given the equations (\ref{wsGSOtwo}), (\ref{GeneralBound}) and the bounds on the spin $h$,  it becomes clear that the world sheet R-charge $q_{\text{ws}}$ can only take the value $
1$. For that value, we have:
\be
q_R^{\bar{a}} = \frac{1}{2} - \frac{2h-1}{k}
\ee and we find the bound:
\be 
|q_R^{\bar{a}}| \le \frac{1}{2} \, 
\ee
on the world sheet Ramond sector R-charge.

To summarize, the left moving Ramond vertex operator for physical states in the covariant formalism, in which the charges are constrained by \eqref{wsGSOtwo} is:
\be
O^{\bar a\, (-\frac12)}_{\text R} = c  ~e^{-\frac{\phi}{2}}
~\Phi_{h}^{w} 
~e^{\ii \sqrt{\frac{2}{k}}(h+\frac{kw}{2}-\frac12) X}
~e^{-\ii(\frac12 +w)\hat H_1 } ~e^{\frac{\ii}{2} \hat H_0}
~\Sigma^{\bar a}~.
\ee
Thus we determined the covariant vertex operators corresponding to the space-time chiral primaries.

\subsubsection{Zero Winding Vertex Operators at a Generic Boundary Point}
In this subsection, we include the spacetime chiral primary operators  into global $sl(2,\mathbb{R})$ multiplets.  We use a variable $x$ to keep track of the different states in one multiplet -- the variable has the intuitive interpretation of designating a point on the boundary of the $AdS_3$ manifold  \cite{Kutasov:1999xu}.
For simplicity, we concentrate on the operators that have  spectral flow number $w=0$ from now on. 
To put the operators $\Phi_h$ of spin $h$ at a generic point $x$, it is useful to temporarily combine a discrete lowest weight representation with a discrete highest weight representation plus a finite dimensional representation, and define the $sl(2)$ multiplet combination $\Phi_h(x)$
\be
\Phi_h(x;z) = \sum_{m\in\mathbb{Z}} \Phi_{h,m,\bar m}(z) x^{-m-h}\bar x^{-\bar m-h} \, \Phi_{h,m,\bar m}(z,\bar z)~.
\ee
Similarly,  we combine the $sl(2,\mathbb{R})$ adjoint fermions \cite{Kutasov:1999xu}
\be
\psi(x;z) = -\psi^+(z) + 2x\psi^3(z)-x^2 \psi^-(z)~.
\ee
 A primary field of spin $s$ in the $sl(2,\mathbb{R})_{k+2}$ model satisfies the   operator product expansion with the currents $j^A(z)$: 
\be
\label{jwithOs}
j^A(z) {\mathcal O}_s(x; z) \approx 
-\frac{D_x^A {\mathcal O}_s(x; z)}{z-w}~,
\ee
where the differential operators representing $sl(2,\mathbb{R})$ read
\be
\label{Dxdefn}
D^-_x = \pa_x~, \qquad D^3_x = x\frac{\pa}{\pa x} + s~,\qquad D^+_x = x^2\pa_x + 2 s x~.
\ee
The fields $\Phi_h(x; z)$ and $\psi(x; z)$ have  spins equal to $h$ and $-1$ respectively. 
The affine $sl(2,\mathbb{R})$ currents also transform in the adjoint:
\be
j(x;z) = -j^+(z) + 2xj^3(z)-x^2 j^-(z)~.
\ee
We can form analogous combinations for the  fermionic current $\hat j(x; z)$ (see Appendix \ref{wstheory} for definitions and conventions) and the supersymmetric current $J(x; z) = j(x;z) + \hat j (x;z)$. The standard operator product expansions between the world sheet fields that we have summarized in Appendix \ref{wstheory} can then be compactly encoded in the  operator product expansions:
\begin{align}
j(x_1;z) \Phi_h (x_2;w) & \approx  \frac{1}{z-w} ((x_1-x_2)^2 \partial_{x_2} - 2 h (x_1-x_2)) \Phi_h(x_2;w) \\
\hat{j}(x_1;z) \psi(x_2;w) & \approx  \frac{1}{z-w} 
((x_1-x_2)^2 \partial_{x_2} +2 (x_1-x_2)) \psi(x_2;w)\\
 \psi(x_1;z_1) \psi(x_2;z_2) & \approx k \frac{(x_{12})^2}{z_{12}}  \, .
\end{align}
In terms of these fields, one can write down the zero spectral flow left-moving vertex operators in the NS sector. We choose a particular normalization whose utility will become apparent:
\begin{align}
\label{vopNS1}
O^{a\, (-1)}_{\text{NS}_1}(x; z) &=\frac{1}{\sqrt{k}(2h-1)}~ c~e^{-\phi}~\Phi_h(x; z)~
e^{\ii \sqrt{\frac{2}{k}} h X}\, V^a_{c.p.}~,~ \quad\quad~ h^a_Y = \frac12-\frac{h}{k}~,&\\
O^{\bar a\, (-1)}_{\text{NS}_2}(x; z)& =\frac{1}{\sqrt{k}}~ c~e^{-\phi}~\Phi_h(x; z)~\psi(x; z)
~e^{\ii \sqrt{\frac{2}{k}}(h -1) X}\, V^{\bar a}_{a.c.p.}~,~~~ h^{\bar a}_Y = \frac{h-1}{k}~.&
\label{vopNS2}
\end{align}
In the Ramond-sector one analogously combines spin fields to form a primary field $S(x)$ such that it is of spin  $s=-\frac12$:
\be
S(x; z) = x\, e^{-\frac{\ii}{2} \hat H_1 + \frac{\ii}{2} \hat H_0} +  e^{\frac{\ii}{2} \hat H_1 - \frac{\ii}{2} \hat H_0}   \, .
\ee
To check that the vertex operator is BRST closed and transforms as claimed,  it is useful to use  the $\hat j^A$ currents expressed in terms of the bosons.
Finally, we have the left-moving Ramond sector vertex operator at a generic point $x$ on the boundary:
\be
O_{\text{R}}^{\bar a\, (-\frac12)}(x;z) = \frac{1}{\sqrt{k}}~ c~e^{-\frac{\phi}{2}}~ \Phi_h(x; z)~ e^{\ii \sqrt{\frac{2}{k}}(h -\frac12) X}~S(x)~\Sigma^{\bar a}~.
\label{RamondHalf}
\ee

\subsection{A Change of Picture}
It will likewise be useful to have the covariant vertex operators available in multiple pictures. We derive the operators in this subsection.
To find the zero picture operators in the NS sector, we apply picture changing to the $(-1)$ picture operators we have found. The latter are schematically of the form:
\begin{equation}
O^{(-1)} = c ~e^{- \phi}~ O_{\text{matter}} 
~,
\end{equation}
where both the matter vertex operator $O_{\text{matter}}$ and $e^{-\phi}$ have dimension equal to $1/2$. 
The vertex operator $O_{matter}$ is a superconformal primary of the ${ N}=1$ world sheet algebra.
The picture changing operator acts on the matter operator as \cite{Friedan:1985ge, Polchinski:1998rr}:
\begin{align}
O^{(0)}_{matter}(0) &= G_{-1/2} \cdot O_{\text{matter}}(0)
 = \oint_0 ~dz~ G(z)~ O_{\text{matter}} (0) 
\, ,
\end{align}
where $G_{-1/2}$ is a mode of the $N=1$ world sheet supercurrent $G$.
The calculation can again be split into three terms corresponding to the factors $AdS_3 \times S^1 \times Y$ -- see equation (\ref{ThreeTermG}). 
It is useful to recall that we suppose $N=2$ superconformal symmetry in the world sheet theory $Y$ and that we have the generic embedding of the  $N=1$ super current $G$ into the $N=2$ superconformal algebra  \cite{Polchinski:1998rr}:
\begin{equation}
G^{Y} = \frac{1}{\sqrt{2}} (G^{+,Y} + G^{-,Y}) \, .
\end{equation} 
Let's implement these statements case by case.

\subsubsection{
The \texorpdfstring{$\text{NS}_1$}{NS1} Case: No Fermion in the AdS Factor}

If there is no  fermion $\psi(x)$ present in the vertex operator, then  picture changing leads to:
\begin{align}
{O}_{\text{NS}_1}^{a\, (0)} &=\frac{1}{\sqrt{k}(2h-1)}~ (\frac{2}{k} \psi_A D^A_x + G^{S^1+Y}_{-1/2} \cdot) \Phi_{h}(x) ~ e^{\ii \sqrt{\frac{2}{k}} h X}~  V^a_{c.p.} 
\, .
\end{align}

\subsubsection{The \texorpdfstring{$\text{NS}_2$}{NS2} Case: A Fermion in the AdS Factor}

We next consider the $NS_2$ case in which the vertex operator \eqref{vopNS2}  has left-moving dimension $H=h-1$. Acting with the picture changing operator we find:
 \begin{align}
 \label{Ozerov1}
 {O}_{\text NS_2}^{\bar a\, (0)} &= \frac{1}{\sqrt{k}}\big(J(x)+ \frac{2}{k} \psi(x) \psi_A D^A_x + \psi(x)\, G^{S^1+Y}\cdot \big) 
 \Phi_{h}(x) ~ e^{\ii \sqrt{\frac{2}{k}}(h -1) X}~ V^{\bar a}_{a.c.p.} ~.
 \end{align}
 The second term in the parentheses can be simplified   :
 \begin{align}
\psi(x)( \psi_A D^A_x)\Phi_h(x) &=(-\psi^+ + 2x\psi^3 - x^2\psi^-)(\frac12 \psi^-D_x^+ + \frac12 \psi^+ D_x^- - \psi^3D_x^3)\Phi_h(x)\cr     
&= \left[\frac{\psi^+\psi^-}{2} (x^2D_x^- - D_x^+) + \psi^3\psi^+(xD_x^- - D_x^3) + \psi^3\psi^-(x D_x^+ - x^2 D_x^3)\right]\Phi_h(x) \cr
&= -\frac{kh}{2} \hat j(x)\Phi_h(x)~.
 \end{align}
 Substituting this into the expression \eqref{Ozerov1} we find 
\begin{align}
 {O}_{\text{NS}_2}^{\bar a\, (0)} &= \frac{1}{\sqrt{k}}~\big(j(x)+(1-h) \hat{j}(x)+ \psi(x) G^{S^1+Y}_{-1/2} \cdot \big) \Phi_{h}(x) ~ e^{\ii \sqrt{\frac{2}{k}}(h -1) X}~ V^{\bar a}_{a.c.p.} ~. 
\end{align}

\subsubsection{Ramond Sector Vertex Operators}
It will be useful to know the Ramond vertex operators  in the $(-\frac32)$ picture as well: 
\be
\label{R3by2operator}
O_{\text{R}}^{\bar a\, (-\frac32)}(x;z) = \frac{2}{ (2h-1)}~c~e^{-\frac{3\phi}{2}}~ \Phi_h(x; z)~ e^{\ii \sqrt{\frac{2}{k}}(h -\frac12) X}~\widetilde S(x)~\Sigma^{\bar a}~,
\ee
where 
\be
\widetilde S(x; z) = - x\, e^{-\frac{\ii}{2} \hat H_1 - \frac{\ii}{2} \hat H_0} +  e^{\frac{\ii}{2} \hat H_1 + \frac{\ii}{2} \hat H_0}~.  
\ee
Acting with the picture changing operator on this operator, we indeed obtain the Ramond operator (\ref{RamondHalf}) in the $(-\frac12)$ picture. The operator $\tilde{S}$ has opposite chirality from the operator $S$.

In summary, we have determined the full spectrum of chiral primaries in space-time. We have described all the relevant covariant vertex operators. For those that are in the zero spectral flow sector, we have provided additional detail. A summary of the latter vertex operators is provided in Table \ref{TheTableOfVertexOperators}.
\begin{table}
\begin{center}
\renewcommand{\arraystretch}{2} 
\begin{tabular}{ |c|c|c| } 
\hline
& Vertex Operator & Dimension $H$ \\
 \hline \hline
$O^{a\, (-1)}_{\text{NS}_1}$ & $\frac{1}{\sqrt{k}(2h-1)}~ c ~  e^{-\phi}~ \Phi_{h}(x) ~
~e^{\ii \sqrt{\frac{2}{k}} h  X}
~V^a_{c.p.} $ & $h$\\
\hline 
 ${O}^{a\, (0)}_{\text{NS}_1} $ & $\frac{1}{\sqrt{k}(2h-1)}~ (\frac{2}{k} \psi_A D^A_x + G^{S^1+Y}_{-1/2} \cdot) \Phi_{h}(x) ~ e^{\ii \sqrt{\frac{2}{k}} h  X}~  V^a_{c.p.} $ & $h$ \\ 
 \hline \hline 
$O^{\bar a\, (-1)}_{\text{NS}_2}$  & $\frac{1}{\sqrt{k}}~c~ e^{-\phi}~ \Phi_{h}(x) ~\psi(x)
~e^{\ii \sqrt{\frac{2}{k}}(h - 1) X} 
 ~V^{\bar a}_{a.c.p.}~ $ & $h-1$ \\
\hline
$O^{\bar a\, (0)}_{\text{NS}_2}$   &$\frac{1}{\sqrt{k}}~ \big(j(x)+(1-h) \hat{j}(x)+ \psi(x) G^{S^1+Y}_{-1/2} \cdot \big) \Phi_{h}(x) ~ e^{\ii \sqrt{\frac{2}{k}}(h - 1) X} ~ V^{\bar a}_{a.c.p.} ~$ & $h-1$ \\
 \hline \hline 
$ O^{\bar a\, (-\frac12)}_{\text R}$ & $\frac{1}{\sqrt{k}}~c  ~ e^{-\frac{\phi}{2}}
\Phi_{h}(x)
~e^{\ii \sqrt{\frac{2}{k}}(h-\frac12) X}
S(x)\, 
\Sigma^{\bar a}~$ & $ h -\frac12 $
\\
\hline
$ O^{\bar a\, (-\frac32)}_{\text R}$ & $\frac{2}{ (2h-1)}~c  ~ e^{-\frac{3\phi}{2}}
\Phi_{h}(x)
~e^{\ii \sqrt{\frac{2}{k}}(h-\frac12) X}
\widetilde S(x)
\Sigma^{\bar a}~$ & $ h -\frac12 $
\\
\hline
\end{tabular}
\end{center}
\caption{The world sheet vertex operators representing space-time chiral primaries and their space-time left-moving conformal dimension, in the subsector with spectral flow number equal to zero.}
\label{TheTableOfVertexOperators}
\end{table}

\section{Chiral Ring Structure Constants}
\label{SpaceTimeChiralRing}
Our goal in this section is to compute  structure constants of the spacetime ${\cal N}=2$ superconformal  chiral ring, involving operators which are in the world sheet sector of zero spectral flow. 
We will derive a  space-time operator product expansion from a world sheet operator product expansion. The advantage of this method is that it only involves data that survives a topological twist of the theory. In particular, we bypass the calculation of the physical three-point functions and two-point functions both in the space-time and on the world sheet. Our method simplifies  calculations in the literature for ${\cal N}=4$ superconformal theories. We also  extend them to the class of ${\cal N}=2$  theories which  exhibit  qualitatively new features.

\subsection{Pictures and Products}
In the NS sector we have determined covariant vertex  operators in  the $(-1)$ as well as the $(0)$ pictures. We wish to calculate the space-time chiral ring. Since the picture number adds in the product of operators, we can multiply an integrated zero picture vertex operator into a fixed $(-1)$ picture vertex operator and obtain once more a $(-1)$ picture operator. Multiplying two vertex operators in the Ramond sector with picture number $(-\frac12)$ leads to an NS operator with picture number minus one. In the R-NS mixed operator product expansions we take the Ramond and NS operators in their canonical pictures to generate a Ramond sector operator in the $(-\frac32)$ picture.

The product of two space-time chiral primary operators inserted at boundary points $x_1$ and $x_2$ is regular as a function of the space-time distance $x_1-x_2$ between the operators. The structure constants of the chiral ring arise from the  zeroth order term in $x_1-x_2$. For example, starting with two operators of the type NS$_2$ we obtain
\begin{multline}
\lim_{x_1 \rightarrow x_2} \int d^2 z_1  O_{\text{NS}_2}^{ \bar a\, (0)}(x_1,\bar{x}_1; z_1,\bar{z}_1)
 O^{\bar b\, (-1)}_{\text{NS}_2} (x_2,\bar{x}_2; z_2,\bar{z}_2) =
 {{\cal A}^{2\bar a\, 2\bar b}}_{2\bar c}
 O^{\bar c\, (-1)}_{\text{NS}_2}(x_2, \bar{x}_2;z_2,\bar{z}_2) \\
   +O(x_1-x_2,\bar{x}_1-\bar{x}_2) \, ,
\label{NSiNSj} 
\end{multline}
where ${\cal A}$ denotes the space-time structure constant for the product under consideration and we have for now only written down one (out of two possible) term(s) on the right hand side. 
We imagine that this calculation takes place inside a perturbative string correlation function. We assume throughout that the operator product in space-time arises from the region in which the world sheet vertex operators are close to each other. See e.g. \cite{Kutasov:1999xu,Aharony:2007rq} for discussions of these points in more generic contexts.

We make one more preliminary remark. We note that we have a graded ring. The Neveu-Schwarz sector vertex operators are even while Ramond sector vertex operators are odd. This  leads to a space-time fermion number grading of the space-time ring.

\subsection{The \texorpdfstring{NS$_2$-NS$_2$ to NS$_2$}{NS2-NS2 to NS2} Structure Constants}
\label{NS222}
In the following subsections, we
 write out our laundry list of structure constants. We start out with the structure constants obtained by  colliding the NS$_2$ operators and giving rise to another NS$_2$ operator as  in equation \eqref{NSiNSj}. We have $H_i=h_i-1$ for  all three operators involved in the operator product expansion. 
For a while, we focus our attention on the world sheet holomorphic sector.  An important ingredient in our calculation is  the operator product expansion between the zero and minus one picture operators:
\begin{multline}
\label{NS-NSOzeroOminusone}
\frac{1}{\sqrt{k}}\int d^2z\bigg(j(x_1; z)+(1-h_1) \hat{j}(x_1; z)+ \psi(x_1; z) G^{S^1+Y}_{-1/2} \cdot\bigg) \Phi_{h_1}(x_1; z)  e^{\ii \sqrt{\frac{2}{k}}(h_1 - 1) X(z)}  V^{\bar a}_{a.c.p.}(z)
\cr
\times \frac{1}{\sqrt{k}}~c~e^{-\phi} ~\psi(x_2; w)~\Phi_{h_2}(x_2; w)~ 
e^{\ii \sqrt{\frac{2}{k}}(h_2 - 1) X(w)} ~ V^{\bar b}_{a.c.p.}(w) 
\, .
\end{multline} 
The first operator is inserted at the world sheet point $z$ while the second operator is inserted at $w$. Before we proceed, let us make a remark about world sheet operator dimensions.
Consider the NS-sector operators at ghost number $-1$. They are of the form  $O^{(-1)}_{\text{NS}}=c\,  e^{-\phi}\, O_{\text{matter}}^{(-1)}$. The total operator has dimension zero. The operator $O_{\text{matter}}^{(-1)}$ has dimension one half. Taking into account the dimension of the fermion operator this leads to the conclusion that the combination 
\be
W_i^{\bar a}(x_i; z) =\Phi_{h_i}(x_i; z)\, e^{\ii \sqrt{\frac{2}{k}}(h_i-1) X}\,  V^{\bar a}_{a.c.p.}(z) ~,
\ee
has dimension zero. Similarly, in ${O}^{(0)}_{\text{matter}}$, the combination $W_i^{\bar a}$  has dimension zero as well since the total operator has  dimension one which equals the dimension of the current.\footnote{A similar reasoning holds in the R-sector.} This will play a role in simplifying our analysis. Moreover, world sheet fermion number conservation will imply important selection rules on our calculations -- we will exploit them in due course. 

We have only shown the left moving contribution to the operator product and we  need to specify the choice of vertex operators in the right moving sector as well. For simplicity, we concentrate on the chiral-chiral ring in space-time, such that we can work with the same type of vertex operators for the space-time right-movers.  One can still have either of the classes of states NS$_{i}$ with $i=1,2$ for the right moving sector. We choose the NS$_{2}$ states also in the right moving sector. From the form of the ghost number zero operator in \eqref{NS-NSOzeroOminusone} we see that there are three independent calculations to be done; schematically the complete contribution to the operator product expansion therefore includes a sum of terms that arises from the   product of left and right moving contributions:
\be
(T_1+T_2+T_3)(\widetilde T_1+\widetilde T_2+\widetilde T_3) \, .
\label{Tsplit}
\ee
We shall evaluate each of these three terms in turn working in the left moving sector first.

 \subsubsection{The First Term}
 \label{NS2NS2firstterm}
 
We begin with the first term that includes the bosonic $AdS_3$ current $j(x;z)$. We need to evaluate the operator product:
\be
\frac{1}{k}~ j(x_1; z) W_{1}^{\bar a}(x_1; z)
\,  \, \, c(w)~ e^{-\phi(w)} ~\psi(x_2; w)~ W_{2}^{\bar b}(x_2; w) ~.
\ee
The fermion, the ghost and the superghosts go along for the ride and we see that if  any space-time chiral primary appears in this operator product at all, it would be an NS$_2$ vertex operator. The only non-trivial factor is the (holomorphic) operator product expansion of the operator $(j \, W_1^{\bar a})(x_1; z)$ with $W_2(x_2; w)$, which we denote by $t_1$. We deal with an operator product expansion of composite operators of the form $(BC)(z)$ with $A(w)$, where we identify $A(w) =W_2^{\bar b}(x_2; w) $, $(B C)(z) = (j \, W_1^{\bar a})(x_1; z)$. What we shall rather do is calculate the operator product expansion of $A(z)$ with $(BC)(w)$ given by the  
generalized Wick theorem for interacting fields \cite{DiFrancesco:1997nk}:
\begin{eqnarray}
\label{genWick}
A(z) (BC)(w) & \approx & 
\frac{1}{2 \pi i } \oint_w  \frac{dy}{y-w} \left( \contraction[2ex]{}{A}{(z)}{B} A(z)B(y) \, \, \,  C(w)+ B(y) \, \, \, 
\contraction[2ex]{}{A}{(z)}{C}
A(z) C(w) \right) \, .
\end{eqnarray} 
To obtain the required operator product expansion we then  interchange $z\leftrightarrow w$, and  Taylor expand the fields that are evaluated at $z$ around the point $w$. We thus first compute:
\begin{align}
t_1
& = \frac{1}{2 \pi i\, k } \oint_w  \frac{dy}{y-w} \left( \contraction[1.4ex]{}{W}{_2^{\bar b}(x_1; z)}{j} W_2^{\bar b}(x_2; z)\,  j(x_1; y)~~W_1^{\bar a}(x_1; w) + j(x_1; y) \, \, \, 
\contraction[1.4ex]{}{W}{_2^{\bar b}(x_2; z)}{W}W_2^{\bar b}(x_2; z) W_1^{\bar a}(x_1; w) \right) \, , 
\end{align} 
We use the basic $\Phi \Phi$ as well as the $j \Phi $ operator product expansions, which are given in the Appendix (equations \eqref{phiphireversed} and \eqref{phijreversed}).  
 Combining these with the operator product expansion in the circle direction and using the definition of the world sheet chiral ring of the $Y$-theory,
\begin{align}
    V^{\bar a}_{a.c.p.}(z)V^{\bar b}_{a.c.p.}(w) &\approx {R^{\bar a\bar b}}_{\bar c} V^{\bar c}_{a.c.p.}(w)~,
\end{align}
one can write down the relevant $WW$ and $Wj$ operator product expansions:
 \begin{align}
 W_2^{\bar b}(x_2; z)W_1^{\bar a}(x_1; w)&\approx  \int d h \,  (w-z)^{\frac{-h(h-1)}{k}+\frac{(h_1+h_2-2)^2}{k}+h^Y_c} 
(x_1-x_2)^{h-h_1-h_2} {C_{h_1,h_2}}^h\, {R^{\bar a \bar b}}_{\bar c} \cr
&\hspace{2cm} \Phi_{h}(x_2;w)\, e^{\ii \sqrt{\frac{2}{k}}(h_1+h_2-2) X } \, V^{\bar c}_{a.c.p.}(w)\\
W_{2}^{\bar b} (x_2;z) j(x_1;y)  & \approx  -\frac{1}{z-y} ((x_2-x_1)^2 \partial_{x_1} - 2 h_2 (x_1-x_2)) W_2^{\bar b}(x_2;z) \cr
&\hspace{3cm}+((x_2-x_1)^2 \partial_{x_1} - 2 h_2 (x_1-x_2)) \partial_z W_2^{\bar  b}(x_2;z)
 \end{align} %
 In the second operator product  we have expanded the $W_2$ operator at $z$. We omit the term proportional to the derivative of the operator as that is a world sheet descendant and  we know that it cannot  lead to a spacetime chiral primary. 
 Substituting into the generalized Wick theorem  \eqref{genWick} leads to the   result:
\begin{align}
t_1  & \approx  \frac{1}{2 \pi i\, k } \oint_w  \frac{dy}{y-w} \Big[ - \frac{1}{z-y} ((x_1-x_2)^2 \partial_{x_2} - 2 h_2 (x_1-x_2)) \contraction[1.4ex]{}{W}{_2^{\bar b}(x_2; z)}{W}W_2^{\bar b}(x_2; z) W_1^{\bar a}(x_1; w) \nonumber \\
&  +   \int d h 
\frac{(x_1-x_2)^{h-h_1-h_2} {C_{h_1h_2}}^h\, {R^{\bar a \bar b}}_{\bar c}}{(w-z)^{\frac{h(h-1)}{k}-\frac{(h_1+h_2-2)^2}{k}-h^Y_c}}  e^{\ii \sqrt{\frac{2}{k}}(h_1+h_2-2) X } \, V^{\bar c}_{a.c.p.}(w)
 \contraction[1.4ex]{}{j}{(x_1; y)}{\Phi} j(x_1; y)\Phi_{h}(x_2;w)
 \Big] \,.
 \end{align}
 Substituting the basic operator product expansions once again we find
 \begin{align}
 t_1 & \approx  \frac{1}{2 \pi i\, k } \oint_w  \frac{dy}{y-w} \Big[ - \frac{1}{z-y} ((x_1-x_2)^2 \partial_{x_2} - 2 h_2 (x_1-x_2))
\nonumber \\
&
\hspace{3cm}\int d h \frac{(x_1-x_2)^{h-h_1-h_2} {C_{h_1 h_2}}^h {R^{\bar a \bar b}}_{\bar c} } 
{(z-w)^{\frac{h(h-1)}{k}-\frac{(h_1+h_2-2)^2}{k} - h_c^Y }}
\, e^{\ii \sqrt{\frac{2}{k}}(h_1+h_2-2) X } V^{\bar c}_{a.c.p.}(w)\Phi_h(x_2; w)  \nonumber \\
&  + \int d h 
\frac{(x_1-x_2)^{h-h_1-h_2} {C_{h_1h_2}}^h\, {R^{\bar a\bar b}}_{\bar c}}{(w-z)^{\frac{h(h-1)}{k}-\frac{(h_1+h_2-2)^2}{k}-h^Y_c} (y-w)}  e^{\ii \sqrt{\frac{2}{k}}(h_1+h_2-2) X } \, V^{\bar c}_{a.c.p.}(w)\cr
&\hspace{7cm}\times ((x_1-x_2)^2 \partial_{x_2} - 2 h (x_1-x_2)) \Phi_h(x_2;w) 
\nonumber \\
&  + \int d h 
\frac{(x_1-x_2)^{h-h_1-h_2} {C_{h_1h_2}}^h\, {R^{\bar a\bar b}}_{\bar c}}{(w-z)^{\frac{h(h-1)}{k}-\frac{(h_1+h_2-2)^2}{k}-h^Y_c} }  e^{\ii \sqrt{\frac{2}{k}}(h_1+h_2-2) X } \, V^{\bar c}_{a.c.p.}(w) \, : j_1(w) \Phi_{h}(x_2;w): 
\Big]~.
\end{align}
By evaluating the integral over $y$, we see that the second term vanishes.  The last term does not give rise to a space-time chiral primary, and so we are left with the first term only:
\begin{align}
 t_1 & \approx  \frac{1}{2 \pi i \, k} \oint_w  \frac{dy}{y-w} \Big[ - \frac{1}{z-y} ((x_1-x_2)^2 \partial_{x_2} - 2 h_2 (x_1-x_2))
\nonumber \\
&
\hspace{1.5cm}\int d h \frac{(x_1-x_2)^{h-h_1-h_2} {C_{h_1 h_2}}^j {R^{\bar a\bar b}}_{\bar c} } 
{(z-w)^{\frac{h(h-1)}{k}-\frac{(h_1+h_2-2)^2}{k} - h_c^Y }}
\, e^{\ii \sqrt{\frac{2}{k}}(h_1+h_2-2) X } V^{\bar c}_{a.c.p.}(w)\Phi_h(x_2; w) \Big]
\end{align}
The $y$-integral simply sets $y=w$. To proceed further we must include the contribution from the right-movers. An important point  is that there is only a single, common integral over the $sl(2,\mathbb{R})$ spin $h$. Reverting to our notation for the operator product in \eqref{Tsplit} we obtain  
\begin{align}
T_1\, \widetilde T_1 \approx \frac{1}{k^2}\int  \frac{dh}{|z-w|^{2+2\frac{h(h-1)}{k}-2\frac{(h_1+h_2-2)^2}{k} - 2h_c^Y }}  {C_{h_1h_2}}^h {R^{\bar a\bar b}}_{\bar c} e^{\ii \sqrt{\frac{2}{k}}(h_1+h_2-2) X } \, V^{\bar c}_{a.c.p.}(w,\bar w) \cr 
\left| ((x_1-x_2)^2 \partial_{x_2} - 2 h_2 (x_1-x_2))\right|^2 |x_1-x_2|^{2h-2h_1-2h_2}\Phi_h(x_2; w, \bar w)
  \label{T1interim}
\end{align}
The structure constants ${C_{h_1 h_2}}^h$ and ${R^{\bar a\bar b}}_{\bar c}$ now include the contribution from the right movers as well.\footnote{To avoid clutter, we do not write out the in principle necessary zoo of indices.} We use once more the fact that we are calculating the  spacetime chiral ring. Spacetime R-charge conservation and unitarity imply \cite{Lerche:1989uy} that  the right hand side of equation \eqref{T1interim} should be  independent of the boundary distance $x_1-x_2$. This  picks out the value  $h=h_3=h_1+h_2-1$ as the one on which to concentrate. The world sheet conformal dimension $h_Y^c$ of the operator $V^c$ is fixed by world sheet $R$-charge conservation to be 
\begin{align}
h^c_Y &= h^a_Y + h^b_Y = \frac{h_1+h_2-2}{k}= \frac{h_3-1}{k}~.
\end{align}
In the second equality we have used the relation between the dimension $h_Y$ and the $sl(2,\mathbb{R})$ spin for the NS$_{2,a}$ vertex operators.  
Combining this with the momentum along the $S^1$ direction we see that this leads to the propagation of an on-shell state that again corresponds to a spacetime chiral primary. This is  consistent with the general discussion of deriving spacetime operator product expansions from world sheet operator product expansions in \cite{Aharony:2007rq} in which the dominant contribution to the $z$-integral arises from on-shell states. To recapitulate, precisely at the value of $h=h_3=h_1+h_2-1$, the  state (whose vertex operator includes the $V^{\bar c}_{a.c.p.}$, $\Phi_h$ and the $S^1$ vertex operator with momentum $p_L = h_3-1$) turns out to be on-shell and this leads to a pole in the $h$-integral. The contribution of this state is  extracted by taking the residue in the $h$-integral\footnote{
Note that we evaluate a pole that is associated to the $|z-w|$ dependence and it therefore differs from the poles in the structure constants, or the poles in the $|x_1-x_2|$ dependence evaluated in \cite{Teschner:1999ug}.}
\be
\text{Res}_{h=h_3} \frac{1}{|z|^{2+f(h)}} = \frac{1}{f'(h_3)}\delta^2(z,\bar z)~,
\ee
where, in our case we have $f'(h_3) = (2h_3-1)/k = (2h_1+2h_2-3)/k$. Performing the $h$-integral  we obtain
\begin{align}
T_1\, \widetilde T_1
  & \approx   \frac{ {C_{h_2,h_1}}^{h_1+h_2-1} {R^{\bar a\bar b}}_{\bar c}}{k(2h_1+2h_2-3)}  \Big|  (x_1-x_2)^2 \partial_{x_2} - 2 h_2 (x_1-x_2)\Big|^2
|x_1-x_2|^{-2} ~ \Phi_{h_1+h_2-1}(w, \bar w)\cr
&\hspace{6cm}
e^{\ii \sqrt{\frac{2}{k}}(h_1+h_2-2) X }~ V^{\bar c}(w,\bar w)\delta^{(2)}(z-w)
\nonumber \\
  & \approx    
  \frac{R^{\bar a\bar b}_c}{k(2h_1+2h_2-3)}
  (1-2 h_2)^2   W^{\bar c}_{h_1+h_2-1}(w, \bar w) \delta^{(2)}(z-w) \, .
\end{align} 
In the second line we have made use of the fact that ${C_{h_2,h_1}}^{h_1+h_2-1} =1$ (see equation (\ref{AStructureConstant})). This takes care of the first  out of nine terms. Given this analysis, the rest of the calculation can be understood swiftly.  
\subsubsection{The Second Term}
We  consider the contribution from the second term in \eqref{NS-NSOzeroOminusone} involving the fermion current $\hat{j}$:
\be
\frac{1}{k}(1-h_1) \hat j(x_1; z) W_{1}^{\bar a}(x_1; z)
\cdot c(w)~ e^{-\phi(w)} ~\psi(x_2; w)~ W_{2}^{\bar b}(x_2; w) ~.
\ee
The non-trivial operator product expansion  involves $\hat j(x_1; z)$ with the fermion $\psi(x_2; w)$ and the operators $W^{\bar a}_1(x_1; z)$ and $W^{\bar b}_2(x_2; z)$.  
We neglect the $(x_1-x_2) \partial_{x_2} \psi(x_2;w)$ term on the right hand side as it does not lead to a chiral ring element. The rest of the analysis parallels the one we already did in detail in the previous subsection. Once again, we have to pair with the right movers and we  obtain the   result:
\begin{align}
T_2 \widetilde T_2 &\approx    
  \frac{ {R^{\bar a\bar b}}_{\bar c}}{k(2h_1+2h_2-3)}
~  (2-2h_1)^2 ~  W^{\bar c}_{h_1+h_2-1}(w, \bar w) \delta^{(2)}(z-w) \, .
\end{align}

\subsubsection{The Result}

The third term $T_3$ in \eqref{NS-NSOzeroOminusone} involves the action of the supercharge $G_{-\frac12}^{S^1+Y}$in the compact directions. Such a term cannot lead to  NS$_2$ operators because of world sheet fermion number conservation -- we shall discuss the action of this operator in subsection \ref{NS2NS2NS1}.
Here, we combine the two contributions $(T_1 +T_2)$ from the left and right movers. Each factor of $T_1$ or $\widetilde T_1$ leads to a factor $(1-2h_2)$ while each factor of $T_2$ or $\widetilde T_2$ leads to the factor $(2-2h_1)$. Taking appropriate linear combinations leads us to:
\begin{align}
(T_1+T_2)(\widetilde T_1+\widetilde T_2) & \approx    
 (2h_1+2h_2-3) \, {R^{\bar a\bar b}}_{\bar c}\, \frac{1}{k}~ W^{\bar c}_{h_1+h_2-1}(w, \bar w) 
  \delta^{(2)}(z-w)\, .
\end{align}
Integrating over $z$ and recalling the other factor operators as well as the R-charge conservation equation $h_3=h_1+h_2-1$, we find the structure constant:
\begin{equation}
{{\cal A}^{2\bar a\, 2\bar b}}_{2\bar c} = (2h_3-1)\,   \, {R^{\bar a\bar b}}_{\bar c} ~.
\end{equation}
We iterate the fact that we have chosen the NS$_{2}^{\bar a}$ vertex operators on both the left and right moving sector to derive this structure constant.   
This structure constant agrees with the structure constant computed in \cite{Dabholkar:2007ey} for models with ${\cal N}=4$ superconformal symmetry. The simple dependence of the structure constant on the space-time spin suggests both an interpretation in terms of a symmetric orbifold, as well as a possible non-renormalization theorem for this structure constant. Indeed, we found here that  universal features known for the ${\cal N}=4$ models extend to this structure constant in all ${\cal N}=2$ superconformal backgrounds under consideration. In the following, we study to what extent these features persist for other structure constants.

\subsection{The \texorpdfstring{NS$_2$-NS$_1$}{NS21} Operator Product Expansion}

Next, we consider the  operator product expansion involving NS sector operators of type 1 and  2. We opt to have the NS$_2$ operator in the $(0)$-picture and wish to evaluate the on-shell space-time chiral primary contribution to:
\begin{multline}
\label{NS2NS1OzeroOminusone}
\frac{1}{\sqrt{k}}\int d^2 z\, \bigg(j(x_1; z)+(1-h_1) \hat{j}(x_1; z)+ \psi(x_1; z) G^{S^1+Y}_{-1/2} \cdot\bigg) \Phi_{h_1}(x_1; z)  e^{\ii \sqrt{\frac{2}{k}}(h_1 - 1) X(z)}  V^{\bar a}_{a.c.p.}(z)
\cr
\times\frac{1}{\sqrt{k}\, (2h_2-1)}~ c~e^{-\phi} ~\Phi_{h_2}(x_2; w)~ 
e^{\ii \sqrt{\frac{2}{k}} h_2 X(w)} ~ V^{b}_{c.p.}(w) 
\, ,
\end{multline} 
when the integrated vertex operator approaches the fixed operator.
The calculation is a simpler counterpart to the one in the previous subsection, so we shall be brief. We denote operator combinations as 
\begin{align}
    W^{\bar a}_{1}(x_1; z) &= \Phi_{h_1}(x_1; z)  e^{\ii \sqrt{\frac{2}{k}}(h_1 - 1) X(z)}  V^{\bar a}_{a.c.p.}(z)\\
    U^b_{2}(x_2; w) &=\Phi_{h_2}(x_2; w)~ 
e^{\ii \sqrt{\frac{2}{k}} h_2 X(w)} ~ V^{b}_{c.p.}(w)  ~.
\end{align}
The operator $W^{\bar a}_1$ has world sheet dimension $0$ while the operator $U^a_{2}$ has dimension equal to $\frac12$. The fermionic current as well as the  $N=1$ creation operator $G^{S^1+Y}_{-1/2}$ appearing in the picture $(0)$ operator do not lead to a spacetime chiral primary.  We  concentrate on the action of the bosonic current operator and consider the operator product expansion:
\be
\frac{1}{k(2h_2-1)}~(j\,  W^{\bar a}_{1})(x_1; z)\cdot U^b_{2}(x_2; w) \, . 
\ee
We again use the formula \eqref{genWick} by first exchanging $z\leftrightarrow w$ and then computing the operator product expansion
\begin{align}
& \frac{1}{2 \pi i\, k(2h_2-1)} \oint_w  \frac{dy}{y-w} \left( \contraction[1.4ex]{}{U}{_2^{b}(x_1; z)}{j} U_2^{b}(x_2; z)\,  j(x_1; y)~~W_1^{\bar a}(x_1; w) + j(x_1; y) \, \, \, 
\contraction[1.4ex]{}{U}{_2^{b}(x_2; z)}{W}U_2^{b}(x_2; z) W_1^{\bar a}(x_1; w) \right) \, .
\end{align}  
To carry out the indicated contractions we peruse the basic $\Phi \Phi$ and the $j \Phi $ operator product expansions in the $AdS_3$ sector.  In addition we also need the operator product expansion in the circle direction and the chiral-anti-chiral operator product expansion in the $Y$-theory. In the first instance, we assume that the sum of the R-charges of the $Y$-vertex operators add up to a positive number. In such a scenario we have (see formula \eqref{CpAcpcp} in the Appendix)
\begin{align}
V_{c.p.}^b(z) V_{a.c.p.}^{\bar{a}}(w) \approx \frac{ {D^{b \bar{a}}}_{c}}{(z-w)^{q^b_Y}}~ V_{c.p.}^{c}(w)~,
\end{align}
where the tensor ${D^{b\bar a}}_c = g^{b \bar d}~ {R^{\bar a \bar e}}_{\bar d} ~ g_{c \bar e}$ is expressed in terms of the chiral ring coefficients $R$ and the Zamolodchikov metric $g$ of the theory $Y$ -- see Appendix \ref{TheoryY}.
The conformal dimension of the world sheet chiral primary $V_{c.p.}^c$ is half of its R-charge, which in turn is the sum of the R-charges of the operators that are multiplied:
\begin{align}
\label{Vccharge}
q_Y^c &= q^{\bar a}_Y + q^b_Y =-2 h^{\bar a}_Y + 2h^b_Y \cr
&= -1+\frac{2}{k}(h_1+h_2-1)~.
\end{align}
The basic operator product expansions that are needed to calculate the product are:
\begin{align}
       U_2^{b}(x_2; z) W_1^{\bar a}(x_1; w) &\approx  \int d h \,   
\frac{(x_1-x_2)^{h-h_1-h_2}}{(w-z)^{\frac12+\frac{h(h-1)}{k}-\frac{(h_1+h_2-1)^2}{k} - h^c_Y}}  {C_{h_1,h_2}}^h\, {D^{b\bar a}}_c    \cr
&\hspace{4cm} \Phi_{h}(x_2;w)\, e^{\ii \sqrt{\frac{2}{k}}(h_1+h_2-1) X } ~V^{c}_{c.p.}(w)\\
 U_2^{b}(x_2; z)\,  j(x_1; y) &\approx-\frac{1}{z-y} ((x_2-x_1)^2 \partial_{x_1} - 2 h_2 (x_1-x_2)) U_2^{ b}(x_2;z) \cr
&\hspace{2.5cm}+((x_2-x_1)^2 \partial_{x_1} - 2 h_2 (x_1-x_2)) \partial_z U_2^{  b}(x_2;z) ~.
\end{align}
From here onward the calculation closely parallels the calculation of the term $t_1$ in the NS$_2$-NS$_2$ operator product in subsection \ref{NS2NS2firstterm}. We have expanded the $U_2^b$ field near the point $z$. Once more, the world sheet derivative of that field will not lead to a spacetime chiral ring element and we omit this term in the subsequent analysis. For a non-trivial on-shell spacetime chiral ring element to be generated in this sector, we need the spin $h=h_3=h_1+h_2-1$. Precisely for this value of the spin $h$, we see from equation \eqref{Vccharge} and the left-moving momentum along the circle that we have a propagating on-shell state. Taking into account the contribution from the right-movers, we again evaluate the residue at the on-shell value for the integrand as a function of the spin $h$ and obtain:
\begin{align}
&  \frac{1}{k^2(2h_2-1)^2}~
  \frac{k~ {C_{h_2,h_1}}^{h_1+h_2-1}~ {D^{b\bar a}}_c} {2h_3-1}~
  (2 h_2-1)^2  ~ U^{c}_{h_1+h_2-1}(w, \bar w) \delta^{(2)}(z-w) \, .
\end{align}
We once more use the the value of the structure constant $ {C_{h_2 h_1}}^{h_1+h_2-1}=1$ and absorb a factor of $\big(\sqrt{k}\, (2h_3-1)\big)^{-2}$ to obtain the correctly normalized closed string NS$_1$ operator. We thereby obtain the structure constant:
\begin{equation}
{{\cal A}^{2\bar a\, 1b}}_{1c} = (2h_3-1)~g^{b \bar d}~ {R^{\bar a \bar e}}_{\bar d} ~ g_{c \bar e} ~  .
\end{equation} 

Let us also   show that the product of an NS$_2$ operator with an NS$_1$ operator can not give rise to an NS$_2$ operator. We have already analyzed the bosonic and fermionic currents appearing in the zero picture operator in \eqref{NS2NS1OzeroOminusone}. It remains to  consider the $G^{S^1+Y}$ term: the $S^1$ term in the $N=1$ supercurrent will not contribute since the fermion associated to the tangent space to the circle is absent from the list of spacetime chiral primaries. The $G^Y$ term will also not contribute to the operator product expansion. This follows from  the identity
\begin{equation}
\langle (G_{-1/2}^+ V_{a.c.p.}^{\bar{a}})(z_1) V_{c.p.}^b(z_2) V_{c.p.}^c(z_3) \rangle 
= 0 \,.  \label{Gacc}
\end{equation}
This is a consequence of pulling off the $G^+$ supercurrent across the sphere.\footnote{An alternative in our context is to argue that due to the bound on the world sheet R-charge of the anti-chiral primaries that can appear in our space-time chiral primaries, a non-zero correlator (\ref{Gacc}) would violate R-charge conservation.} The identity  implies that one cannot produce an anti-chiral primary in the operator product expansion of $(G_{-1/2}^+ V_{a.c.p.})(z_1)$ and $V_{c.p.}(z_2)$. Thus, we have proven the vanishing
\be
{{\cal A}^{1\bar a\, 2b}}_{2\bar c} =0~.
\ee

\subsection{The \texorpdfstring{NS$_1$-NS$_1$}{NS11} Operator Product Expansion}

We turn to the remaining operator product expansion in the NS sector involving two operators of type 1. Up to normalization factors we have the operator product:
\begin{multline}
\label{NS1NS1OzeroOminusone}
    \int d^2z~ (\frac{2}{k} \psi_A D^A_{x_1} + G^{S^1+Y}_{-1/2} \cdot) \Phi_{h_1}(x_1; z) ~ e^{\ii \sqrt{\frac{2}{k}} h_1  X}~  V^a_{c.p.} (z)
\\
\times c~e^{-\phi} ~\Phi_{h_2}(x_2; w)~ 
e^{\ii \sqrt{\frac{2}{k}} h_2 X(w)} ~ V^{b}_{c.p.}(w) 
\, .
\end{multline} 
The first term can only potentially give rise to an NS$_2$ vertex operator in the minus one picture as it has an $AdS_3$ fermion. However, note that the resulting vertex operator in the $(-1)$ picture will necessarily have a chiral primary in the $Y$ sector and such an operator is not in the list of chiral primaries. Therefore, we find that
\be
{{\cal A}^{1a\, 1b}}_{2\bar c} = 0~.
\ee

Let us similarly consider the second term in \eqref{NS1NS1OzeroOminusone}; this can potentially give rise to a NS$_1$ vertex operator in the minus one picture. As before, the term $G^{S^1}$ does not contribute. We therefore focus on the $G^Y$ term. In the $Y$ sector, we have an operator product that is determined by the  
three-point function:
\be
\langle (G_{-1/2}^- V_{c.p.}^a)(z_1) V_{c.p.}^b(z_2) V_{a.c.p.}^{\bar{c}}(z_3) \rangle =0\, .
\ee
By utilizing the superconformal Ward identity one can indeed show that  this correlation function vanishes.\footnote{See e.g. \cite{deBoer:2008ss} for the detailed derivation.} Therefore we have the vanishing ${\cal N}=2$ chiral ring structure constant:
\be
{{\cal A}^{1a\, 1b}}_{1 c} = 0~.
\ee
The tallying reader realizes that in the NS sector, there is one remaining structure constant to compute in which NS$_2$ operators produce a NS$_1$ operator. We will discuss this possibility in subsection \ref{NS2NS2NS1}.

\subsection{The \texorpdfstring{NS$_2$}{NS2}-R Operator Product Expansion}

We turn to  operator product expansions involving one NS sector and one Ramond sector vertex operator. We use the integrated Ramond vertex operator in the $(-\frac{1}{2})$ picture and the unintegrated NS$_2$ operator in the $(-1)$ picture. The result is an unintegrated Ramond sector operator in the $(-\frac32)$ picture. We study the operator product: 
\begin{multline}
\frac{1}{\sqrt{k}}\int d^2z~ e^{-\frac{\phi}{2}} \Phi_{h_1}(x_1; z)~S(x_1; z)~ e^{\ii\sqrt{\frac{2}{k}}(h_1-\frac12) X(z) }~\Sigma^{\bar a}\\
\times\frac{1}{\sqrt{k}}~ c~ e^{-\phi}~\Phi_{h_2}(x_2; w)~\psi(x_2; w)~ e^{\ii\sqrt{\frac{2}{k}}\,   (h_2-1)\, X(w) }~ V_{a.c.p.}^{\bar b}(w) \, .
\end{multline}
In addition to the operator product expansions between the $AdS_3$  operators $\Phi_{h_i}$ and the operator product expansions in the circle sector, we use the expansions:
\begin{align}
e^{-\frac{\phi(z)}{2}}\cdot e^{-\phi(w)} &\approx \frac{e^{-\frac{3\phi(w)}{2}}}{(z-w)^{\frac12}}\\
\frac{1}{\sqrt{k}}\psi(x_1;z) S(x_2;w) &\approx 
2(x_1-x_2) \frac{\widetilde{S}(x_2;w)}{(z-w)^{\frac12}}
\label{FermionWithSpinField}\\
V^{\bar b}_{a.c.p.}(z)  \Sigma^{\bar a}~ (w)
&\approx \frac{  {R^{\bar b\bar a}}_{\bar c} \, \Sigma^{\bar c}(w) }{(z-w)^{h^{\bar{b}}_Y}}~.
\end{align}
In the last operator product  we  use the fact that $h_Y^{\bar{b}} = \frac{h_2-1}{k}$. We once again add the right moving contribution and proceed as before. In the limit $x_1\rightarrow x_2$, we find that the leading non-zero contribution arises from the spin $h=h_3= h_1+h_2-1$. This precisely leads to a spacetime chiral primary generated by a world sheet Ramond sector operator in the $(-\frac32)$ picture. The contribution by this on-shell state is obtained from the residue in the $h$-integral and leads to a $\delta$-function that localizes the $z$-integral. We obtain the closed string vertex operator: 
\be
  \frac{4\, {C_{h_1,h_2}}^{h_1+h_2-1} \, {R^{\bar b\bar a}}_c}{(2h_3-1)}~c~\bar c~ e^{-\frac{3\phi}{2}-\frac{3\tilde\phi}{2}}~\Phi_{h_1+h_2-1}(x_2; w, \bar w)~ e^{\ii (h_1+h_2-\frac32) X(w, \bar w)}~\widetilde S(x_2; w, \bar w)~\Sigma^a(w, \bar w)~.
\ee
 By keeping track of the normalization of the $(-\frac32, -\frac32)$ picture operator (see \eqref{R3by2operator}) and using $C^{h_1+h_2-1}_{h_1 h_2}=1$, we obtain the structure constant
\be
{{\cal B}^{2\bar a\, \bar b}}_{\bar c} = (2h_3-1) {R^{\bar b \bar a}}_{\bar c}~.
\ee
If we consider the operator product expansion of the Ramond operator with the NS$_1$ vertex operator we do not find a space-time chiral ring element in the $(-\frac32)$-picture as the spin field retains its chirality.
We conclude that this structure constant is zero:
\be
{{\cal B}^{1 a\,  b}}_{c}=0~.
\ee

\subsection{The R-R Operator Product Expansion}

Finally, we evaluate the operator product expansion involving one integrated Ramond vertex operator with picture number $(-\frac12)$ and an unintegrated Ramond vertex operator with picture number $(-\frac12)$. The left moving contribution is given by
\begin{multline}
\frac{1}{\sqrt{k}}\int d^2z~ e^{-\frac{\phi}{2}} \Phi_{h_1}(x_1; z)~S(x_1; z)~ e^{\ii\sqrt{\frac{2}{k}}(h_1-\frac12) X(z) }~\Sigma^{\bar a}(z)\\
\times\frac{1}{\sqrt{k}}~ c~ e^{-\frac{\phi}{2}}~\Phi_{h_2}(x_2; w)~S(x_2; w)~ e^{\ii\sqrt{\frac{2}{k}}\,   (h_2-\frac12)\, X(w) }~ \Sigma^{\bar b}(w)~.
\end{multline}
The bosonic ghost is transcribed trivially while the operator product expansions in the bosonic $AdS_3 \times S^1$ sectors are the same as before. The new operator product expansions involved in this calculation are
\begin{align}
    e^{-\frac{\phi(z)}{2}}e^{-\frac{\phi(w)}{2}} &\approx \frac{e^{-\phi(w)}}{(z-w)^{\frac14}}\\
    S(x_1; z)S(x_2; w) &\approx -\frac{x_1-x_2}{(z-w)^{\frac12}}\\
\Sigma^{\bar{a}}(z) \Sigma^{\bar{b}} (w) & \approx 
\frac{ {{R}^{\bar a\bar b}}_{\bar c} {M^{\bar c}}_{c} V_{c.p.}^{c} }{(z-w)^{h_Y^c-\frac{c_Y}{12}}}  \, .
\end{align}
Here ${M^{\bar c}}_c$ is the real structure associated to the $N=2$ superconformal theory on $Y$ \cite{Cecotti:1991me}. 
In a second step, we  include once more the contribution of the right-movers and in the $x_1\rightarrow x_2$ limit the leading non-zero term picks out the spin $j=h_1+h_2-1$ in the $\Phi$-$\Phi$ operator product expansion. Regarding the operator product expansion in the $Y$-sector we see that,  provided that the R-charges of the Ramond ground states add up to a positive real number, we obtain a world sheet chiral primary whose conformal dimension is half of its R-charge  (which in turn is determined by the sum of the R-charges of the Ramond ground states). Proceeding as before,  we obtain the operator product expansion coefficient
\begin{equation}
{{\cal C}^{\bar a \bar b}}_{1c} =   (2h_3-1)~{{R}^{\bar a\bar b}}_{\bar c} {M^{\bar c}}_{c}  \, .
\end{equation}
It is clear from the derivation that it is impossible to obtain an $AdS_3$ fermion on the right hand side. Thus the structure constant associated to the R-R to NS$_2$ operator product is zero, ${{
\cal C}^{\bar a \bar b}}_{2\bar c} = 0$. 

\subsection{The Structure Constants from Topological Data}

Let us collect the results we have computed  and simplify them by redefining the class of $\text{NS}_1$ operators using the Zamolodchikov metric: 
\begin{equation}
    O^{(-1)}_{\bar a, \text{NS}_1} = g_{\bar a b}~O^{b\, (-1)}_{\text{NS}_1}~.
\end{equation}
In a more condensed notation, the non-vanishing structure constants of the spacetime chiral ring computed so far are:
\begin{align}
   {\text{NS}_{2}}^{\bar a} \times {\text{NS}_{2}}^{\bar b} &\longrightarrow ~(2h_3-1) ~{R^{\bar a\bar b}}_{\bar c}~ {\text{NS}_{2}}^{\bar c} 
   \label{NS2NS2NS2redef} \\
   {\text{NS}_{2}}^{\bar a} \times {\text{NS}_{1}}_{\bar b} &\longrightarrow ~(2h_3-1)~{R^{\bar a\bar c}}_{\bar b}~ {\text{NS}_{1}}_{\bar c}\\
    {\text{NS}_{2}}^{\bar a} \times 
    \text{Ra}^{\bar b}\hspace{.2cm} &\longrightarrow~(2h_3-1)~ {R^{\bar a\bar b}}_{\bar c}~\text{Ra}^{\bar c}\\
    \text{Ra}^{\bar a} \times \text{Ra}^{\bar b} \hspace{.2cm} &\longrightarrow ~(2h_3-1)~ {R^{\bar a\bar b}}_{\bar c}~ \eta^{\bar c \bar d}~{\text{NS}_{1}}_{\bar d}~, \label{RaRaNS1redef}
\end{align}
where $h_3=h_1+h_2-1$ and $\text{Ra}^{\bar{a}}$ denotes the Ramond sector operators. 
We have defined the topological metric $\eta^{\bar c\bar d} = {M^{\bar c}}_c\, g^{c\bar d}$ which is the two point function of world sheet chiral primaries in the twisted topological theory  $Y$ \cite{Cecotti:1991me}.  
We conclude  that these four classes of structure constants are entirely captured by the world sheet anti-chiral ring of theory $Y$. 

Finally, we recall that our notation is compact. For instance, in equation (\ref{NS2NS2NS2redef}), when we consider a space-time $(c,c)$ ring element, we can combine left- and right moving operators, and would find the index structure:
\begin{align}
   {\text{NS}_{2}}^{\bar{a}_L,\bar{a}_R} \times {\text{NS}_{2}}^{\bar{b}_L,\bar{b}_R} &\longrightarrow ~(2h_3-1) ~{R^{\bar{a}_L \bar{b}_L}}_{\bar{c}_L}~{R^{\bar{a}_R \bar{b}_R}}_{\bar{c}_R}~ {\text{NS}_{2}}^{\bar{c}_L \bar{c}_R} \, ,
   \end{align}
   where the ring structure constants correspond to the $(a,a)$ world sheet ring and the indices need to be chosen such that the left-right world sheet vertex operator occurs in the chosen modular invariant string spectrum. This is but one example of how to unpack our compact notation in equations (\ref{NS2NS2NS2redef}) to (\ref{RaRaNS1redef}).

\subsection{The \texorpdfstring{NS$_2$-NS$_2$ to NS$_1$}{NS22to1} Structure Constants}
\label{FourToTwo}
\label{NS2NS2NS1}
We are ready to study the odd duckling left over from the NS$_2$-NS$_2$ operator product expansion.
The NS$_2$-NS$_2$ into NS$_1$ structure constant is the only case in which we have to  compute the action of the supercurrent on the operators in the theory $Y$ in order to compute the space-time chiral ring relation.  Therefore, in this case, the space-time structure constant depends not only on the quantum numbers and world sheet chiral ring operators, but also on the three-point function of a world sheet superconformal descendant. To compute the structure constant, we consider explicit models. A first model is the counterpart to the ${\cal N}=4$ superconformal model $AdS_3 \times S^3$ -- we consider the T-dual background $AdS_3 \times S^1 \times MM \times M^4$ where the minimal model $MM$ has central charge $c_{MM}=3-6/k$ and $M^4$ represents a central charge six world sheet conformal field theory.\footnote{For simplicity we ignore orbifold identifications.} We then generalize our analysis to models $AdS_3 \times S^1 \times \prod_{i=1}^3 MM_i$ where the three minimal model levels satisfy $\sum_{i=1}^3 k_i^{-1}=k^{-1}$ for criticality. 

\subsubsection{The \texorpdfstring{${\cal N}=4$}{N=4} Model Revisited}

The calculation of the structure constant proceeds as in previous subsections. We begin with the operator product 
\begin{multline}
\label{ns2ns2ns1}
\frac{1}{\sqrt{k}}\int d^2 z\,  \Phi_{h_1}(x_1; z)  e^{\ii \sqrt{\frac{2}{k}}(h_1 - 1) X(z)} ~\psi(x_1; z)~ \left(G^{Y}_{-1/2} \cdot V^{\bar a}_{a.c.p.} \right)(z)
\cr
\times \frac{1}{\sqrt{k}}~c~e^{-\phi} ~\psi(x_2; w)~\Phi_{h_2}(x_2; w)~ 
e^{\ii \sqrt{\frac{2}{k}}(h_2 - 1) X(w)} ~ V^{\bar b}_{a.c.p.}(w) 
\, .
\end{multline} 
The ghost and superghost factors, the bosonic factors and the $AdS_3$ fermions all have operator products  that we have already encountered. We focus on the new operator product that arises in the $Y$-sector.  
We consider a theory $Y=MM\times M^4$ where  $M^4$ is geometrical and therefore  $M^4=K3$ or $M^4=T^4$. These theories have no (anti)-chiral primary of dimension smaller than one half, and therefore only the minimal model theory provides non-trivial operators in the vertex operators describing space-time chiral primaries. 

Our conventions for the minimal model are those of \cite{Maldacena:2001ky}.\footnote{However, we work with a supersymmetric level $k=k_{bos}+2$. } The super parafermion fields are $\psi_{j,n,s}$ where $j$ is the $su(2)$ spin, $n$ is a spin component and $s$ denotes the fermion number. The world sheet conformal dimension and U$(1)_R$-charge of the parafermions are given by
\begin{align}
    h(j,n,s) &= \frac{j(j+1)}{k}- \frac{n^2}{4k} + \frac{s^2}{8}~,\qquad q(j,n,s) = \frac{s}{2}-\frac{n}{k}~.
\end{align}
The anti-chiral primaries that appear in \eqref{ns2ns2ns1} correspond to the fields $\psi_{j,2j,0}$. The $su(2)$ spin of the colliding operators are related to the $AdS_3$ spin $h_i$ by the relation $j_i=h_i-1$. 
From the operator product it is clear that in the $Y$ sector we need  the coupling between the descendant of an anti-chiral field and an anti-chiral primary field. For an NS$_1$ operator to appear as a result, the fusion should give rise to a chiral primary field. The  relevant operator product is given by  
\begin{align}
(G^+_{-1/2} \psi_{j_1,2j_1,0})(z) \psi_{j_2,2j_2,0}(w) & \approx  \frac{ {{C^{su(2)}}_{j_1j_2}}^{\frac{k}{2}-\tilde{j}_3-1} }{(z-w)^{\frac{\tilde{j}_3-j_1-j_2}{k} +\frac{1}{2}}}~    \psi_{\frac{k}{2}-\tilde{j}_3-1,-2 \tilde{j}_3+k,2}(w) +\dots 
\, .
\label{MMOPE}
\end{align}
 In the parafermion variables a chiral primary corresponds to the field $\psi_{j,-2j,0}$.
However there is an equivalence relation $\psi_{j,n,s} \equiv \psi_{k/2-j-1,n+k,s+2}$ which allows us to identify the state on the right hand side of (\ref{MMOPE}) as a chiral primary $\psi_{\tilde{j}_3,-2\tilde{j}_3,0}$.  The action of the operator $G^+_{-1/2}$ on the anti-chiral primary is to augment the $s$ quantum number by two. Moreover, by R-charge conservation we obtain 
\be
1-\frac{2j_1+2j_2}{k} = \frac{2\tilde{j}_3}{k}~.
\ee
We  define the spin $j_3 = k/2-\tilde{j}_3-1$  in terms of which we find the simpler relation $j_3=j_1+j_2-1$.
 Furthermore, the conformal dimension of the chiral primary field in terms of the new variable is given by
\be
\label{cpdimension}
h_{c.p.} = \frac12 - \frac{j_3+1}{k} = \frac12 - \frac{j_1+j_2}{k}~.
\ee
To summarize: from all possible fusions, we pick the one labelled by the spin $\tilde{j}_3$ as it leads to a chiral primary in the minimal model, which is crucial to obtain an NS$_1$ operator on the right hand side. The parafermion structure constant reduces to an $su(2)$ Wess-Zumino-Witten model structure constant \cite{Zamolodchikov:1986gh}.
The relevant $su(2)$ structure constant is recorded in equation (\ref{SU(2)StructureConstant}) in Appendix \ref{WS}. Given the fusion in the compact sector, we proceed as before and take the limit $x_1\rightarrow x_2$. The leading non-zero contribution arises when the intermediate $AdS_3$ spin equals $h_3 = h_1+h_2-2 = j_1+j_2$, where we have used the relation between the spins in the $AdS_3$ and the compact sectors. Comparing with equation \eqref{cpdimension} and using the momentum along the $S^1$ direction we see that we have obtained a propagating on-shell state that corresponds to a spacetime chiral primary of the type NS$_1$  (\ref{cpns1defn}). 

We perform a similar analysis for the right-movers and execute the $h$-integral as before. 
We obtain the structure constant \cite{Dabholkar:2007ey}:
\be
{{\cal A}^{1\bar a 2\bar a}}_{3c} = 
(2h_3-1)\, {C_{h_1, h_2}}^{h_1+h_2-2} {{C^{su(2)}}_{j_1 j_2}}^{j_1+j_2-1}~.
\ee
We note that since $h_3=h_1+h_2-2$, unlike the previous cases, we encounter a non-trivial structure constant (\ref{ASecondStructureConstant}) in the $AdS_3$ sector. At the same time we also have a non-trivial structure constant arising from the compact sector. 
As observed in \cite{Dabholkar:2007ey} in the ${\cal N}=4$ supersymmetric models, a key point is that  the $AdS_3$ and $su(2)$ structure constants cancel up to a factor of $\nu^{-1}$, and we are left with the  structure constant proportional to the spins, namely $2h_3-1$. Of course, we obtain a non-zero structure constant only when the fusion is allowed by the $su(2)$ fusion rules
\cite{Dabholkar:2007ey}.

\subsubsection{A Generalization}
It would certainly be interesting to compute the NS$_2$-NS$_2$-NS$_1$ structure constant for an arbitrary choice of theory $Y$. 
To obtain the structure constant, we must evaluate the correlator 
\be
\langle G^{+,Y}_{-1/2} \cdot V_{a.c.p.}^a(z_1) V_{a.c.p.}^b (z_2) V_{a.c.p.}^c(z_3) \rangle
\ee
in theory $Y$. 
When we have a factorized theory $Y$, the $G^{+,Y}$ current is a sum of super currents in the individual factors. Then we must have that the operators in all factors but one are the identity operator. Otherwise world sheet R-charge conservation in the factor not containing the $G^{+}_{-1/2}$ term will set the structure constant to zero. 

We shall settle for remarking how to compute the structure constant in an infinite but restricted class of ${\cal N}=2$ theories. We
 choose the theory $Y$ to be a product of three minimal models at levels $k_{i=1,2,3}$. This represents  a large class of models since we can allow any levels $k_i$ for the three minimal models as long as we choose $ k^{-1}=\sum_{i=1}^3 k_i^{-1} $. The operator $G^{+,Y}_{-1/2}$ in the theory $Y$ becomes a sum of operators, each non-trivial in a given factor:
\begin{align}
G^{+,Y}_{-1/2} &=G^{+,MM_1}_{-1/2} + G^{+,MM_2}_{-1/2}+ G^{+,MM_3}_{-1/2}
\, .
\end{align}
As we argued, a set of chiral primary operators with non-zero structure constants must correspond to the identity operators in all but one minimal model. We label the single non-trivial minimal model factor that enters a particular calculation by an index $l$. The on-shell constraints  for the operators in the structure constant calculation read:
\begin{align}
\text{NS}_1:&  \quad
\frac{\tilde{j}_3}{k_l} = \frac{1}{2}-\frac{h_3}{k} =
\frac{1}{2} -
\frac{{j}_3+1}{k_l}
 \nonumber \\
 \text{NS}_2:& \quad
 \frac{j_{1,2}}{k_l}=\frac{h_{1,2}-1}{k} \, .
 \label{jvshN=2}
\end{align}
Space-time R-charge conservation still implies that $h_3=h_1+h_2-2$ while world sheet R-charge conservation enforces $j_3=j_1+j_2-1$.  
The structure constant is computed as in the previous subsection, and reads:
\be
{{\cal A}^{1\bar a 2\bar a}}_{3c} (l) = 
 (2h_3-1) {C_{h_1, h_2}}^{h_1+h_2-2} {{C^{su(2)_{k_l}}}_{j_1 j_2}}^{j_1+j_2-1}~.
\ee
To evaluate the structure constant, we make an important observation. The structure constant for the $AdS_3$ model simplifies drastically and generically when the space-time R-charge constraint is taken into account:\footnote{This is independent of  whether $2h_i-1$ is an integer.}
\begin{align}
{C_{h_1, h_2}}^{h_1+h_2-2} {{C^{su(2)_{k_l}}}_{j_1 j_2}}^{j_1+j_2-1} &=\nu^{-1}  \,  \frac{\gamma(\frac{2h_1-2}{k})\gamma(\frac{2h_2-2}{k}) }{
\gamma(\frac{2h_1+2h_2-4}{k})}
\times \frac{\gamma(\frac{2j_1+2j_2}{k_l})}{
\gamma(\frac{2j_1}{k_l}) \gamma(\frac{2j_2}{k_l})}
= \nu^{-1} \, .
\end{align}
Thus, for this large class of ${\cal N}=2$ models, we again find the same structure constant after a still more intricate cancellation. 
It would be  interesting to calculate these structure constants for more general models $Y$ or to prove that the cancellation we observed for our class of models persists in all cases.

\subsection{Additional Observations on the Structure Constants}

We saw that with a convenient normalization of the  vertex operators, four of the structure constants are proportional to  $2h_3-1$ where $h_i$ are the world sheet spins of the operators and $h_3=h_1+h_2-1$. The dependence on the $Y$ theory was through the structure constants of its (anti-) chiral ring. This generalizes  the structure constants in the literature \cite{Dabholkar:2007ey} for the ${\cal N}=4$ superconformal models to a large class of ${\cal N}=2$ superconformal models. Introducing the number $n_i=2h_i-1+kw_i$, we see that the universal part of the structure constant equals $n_1+n_2-1$, which we recognize as a combinatorial structure constant arising in symmetric orbifold conformal field theory, where the numbers $n_i$ have the interpretation as the lengths of permutation cycles. This is described in  \cite{Lunin:2000yv} (and combinatorially in e.g. \cite{Li:2020nei}). 

In the case of ${\cal N}=4$ space-time superconformal symmetry, a non-renormalization theorem was proven that shows that the structure constants of the chiral ring are covariantly constant on the moduli space \cite{deBoer:2008ss}. This allows for a natural matching of structure constants between a bulk NSNS string point and a symmetric orbifold conformal field theory \cite{Dabholkar:2007ey,Gaberdiel:2007vu}. The non-renormalization theorem shows that in ${\cal N}=4$ theories, all marginal deformations can be constructed from anti-chiral primaries, and in these directions, structure constants of the chiral ring are covariantly constant \cite{deBoer:2008ss}. The proof of the theorem makes use of the full ${\cal N}=4$ algebra. 

Our results for ${\cal N}=2$ theories, for which in principle structure constants can depend on marginal deformations built on space-time chiral primaries, beg the question of why they seem similarly universal. Of course, there is a factorized dependence on the world sheet chiral ring structure constants of the theory $Y$ which can vary under space-time marginal deformations, and relevant deformation in $Y$. The combinatorial prefactor $n_1+n_2-1$ however, appears to be universal. 

For the other structure constant we computed, we confirmed that in ${\cal N}=4$ theories, it is proportional to $n_1+n_2-3$, associated to different and interesting combinatorics in the symmetric orbifold \cite{Lunin:2000yv}.\footnote{See also \cite{Li:2020nei} for the relation to a co-product in the topological conformal field theory in the ${\cal N}=4$ context.} Moreover, for a  class of ${\cal N}=2$ superconformal models that arise as a tensor product of minimal models, we showed that the same combinatorial prefactor appears in the structure constant. This raises the question as to whether these properties will continue to be true for a  generic ${\cal N}=2$ model.

\section{Conclusions}
\label{Conclusions}
In this paper, we studied the bulk string theory background $AdS_3 \times S^1 \times Y$ where $Y$ corresponds to a  conformal field theory with $N=2$ superconformal symmetry on the world sheet. This is a class of string theory backgrounds dual to a space-time conformal field theory with ${\cal N}=2$ superconformal symmetry. The background with Neveu-Schwarz-Neveu-Schwarz flux is exactly solvable in the inverse string tension expansion and allows a calculation of the spectrum and correlation functions exactly in the parameter $\alpha'$ divided by the radius of curvature squared. This is a large class of backgrounds with  extended supersymmetry and they therefore have a chiral ring. These backgrounds and observables form a wonderful testing ground to attempt to prove a topological subsector of the anti-de Sitter/conformal field theory correspondence.

Firstly, we fully determined the spectrum of space-time chiral primaries using the exact world sheet description of the string propagating in $AdS_3 \times S^1 \times Y$. Our results classify the  space-time chiral primaries in terms of the chiral primaries, anti-chiral primaries, and Ramond ground states of the $N=2$ superconformal world sheet theory $Y$. An upper bound on the world sheet R-charge of the operators that enter the calculation emerges. It lays bare a close connection between the space-time chiral primaries and supersymmetry preserving relevant deformations of theory $Y$.  Secondly, we  computed a subset of the structure constants of the chiral primary ring. Our method of calculation consisted in determining the leading order in the chiral operator product expansion by exploiting the operator product of the corresponding world sheet vertex operators. The method concentrates on purely topological data in the space-time theory and gives rise to  results that are simple and universal. We applied the general method to all structure constants involving operators with zero spectral flow. 

We demonstrated that the space-time chiral ring structure constants involved the world sheet chiral ring structure constants, the Zamolodchikov metric in the chiral sector as well as the world sheet real structure that relates two canonical bases of Ramond ground states.  When expressed in terms of suitably chosen space-time chiral primaries, we could show that a subset of space-time structure constants only depend on the topological data of the theory $Y$. We also noted a dependence of a structure constant on  a global superconformal descendant. In the ${\cal N}=4$ supersymmetric setting, it was demonstrated that this structure constant is tied to a co-product defined in  terms of the central charge six factor of theory $Y$. It is a definite challenge to extend this algebraic structure to the ${\cal N}=2$ setting.

There is a clear open problem at hand, which is to compute all possible structure constants among the full set of chiral primaries, including all operators with non-zero winding. It is tempting to speculate that the number $n_i=2h_i-1$ appearing in the structure constants will be generalized to $n_i=2h_i-1+kw_i$ for operators including winding, but it is important to demonstrate this explicitly. 

If non-renormalization theorems exist for these backgrounds, they need to be proven. They are obviously crucial in understanding the matching of structure constants at different points in the moduli space, accessible in the bulk or boundary theory.  

There is an intriguing global question to be answered. Clearly, the algebraic $N=2$ structure of the theory $Y$, including its relevant deformations, structure constants and Zamolodchikov metric enter the topological theory in spacetime. Can we express the space-time chiral ring in terms of these world sheet structures ? Our work  gives a partial answer to this question, but more work is needed to characterize the precise relation. The full answer may invoke the topological-anti-topological structure of the theory $Y$, and may involve a symmetric group Frobenius algebra.  Finally, we can ask once more whether there is a reformulation of the space-time theory that renders the calculation of these structure constants even simpler by projecting onto space-time chiral primaries at every step.

\appendix

\appendix

\section {The World Sheet Theory on \texorpdfstring{$AdS_3\times S^1$}{ads3 s1}}
\label{wstheory}
\label{WS}
In this appendix, we gather  data on the world sheet conformal field theories that are necessary to perform the calculations in the bulk of the paper. The $AdS_3 \times S^1 \times Y$ target space gives rise to  largely factorized world sheet conformal field theories. We discuss these in turn. 

\subsection{The Supersymmetric \texorpdfstring{$AdS_3$}{AdS3} Factor}

We define  the affine currents $J^A$ of a supersymmetric $sl(2,\mathbb{R})$ WZW model at level $k$, with operator product expansions\footnote{We follow the notations and conventions of \cite{Giveon:2003ku}.}:
\begin{align}
J^A (z) J^B(w) &\approx  \frac{k}{2} \frac{ \eta^{AB}}{(z-w)^2} + \frac{\ii \epsilon^{ABC}\eta_{CD}J^D(w)}{z-w} \\
J^A(z) \psi^B(w) &\approx \frac{\ii \epsilon^{ABC}\eta_{CD}\psi^D(w)}{z-w}\\
\psi^A(z) \psi^B(w) &\approx \frac{k}{2} \frac{\eta^{AB}}{z-w}~,
\end{align}
with $\eta^{AB} =\text{diag} (1,1,-1)$ and $\epsilon^{123} =1$. One can define a combination of currents $j^A$ such that they form a bosonic $sl(2,\mathbb{R})$ affine algebra at level $k+2$:
\be
j^A = J^A  + \frac{\ii}{k}\epsilon^{ABC}\eta_{BD}\eta_{CE} \psi^D\psi^E~.
\ee
The bosonic currents $j^A$ have regular operator product expansion with the fermions. 
The three fermions $\psi^A$ generate an $sl(2,\mathbb{R})$  algebra at level $-2$. The algebra is generated by the currents
\be
\hat j^A = -\frac{\ii}{k} \epsilon^{ABC}\eta_{BD}\eta_{CE} \psi^D\psi^E~. 
\ee
In component form the currents read: 
\be
\hat j^1 = \frac{2\ii}{k}\psi^2\psi^3~,\quad \hat j^2 = \frac{2\ii}{k} \psi^3\psi^1~,\quad \hat j^3 = -\frac{2\ii}{k} \psi^1\psi^2~.
\ee
The total current $J^A$ is then the sum of the bosonic and fermionic ones: 
\be
J^A = j^A + \hat j^A~.
\ee
We transform the fermions and the currents into a light cone $(+,-,3)$ basis by defining the combinations
\be
\psi^{\pm} = \psi^1 \pm \ii \psi^2~, \quad \hat j^{\pm} = \hat j^1\pm \ii \hat j^2~. 
\ee
Then the components are
\be
\hat j^+ = -\frac{2}{k}\psi^3 \psi^+~,\quad\hat j^3 = \frac{1}{k}\psi^+\psi^-~, \qquad  \hat j^- = \frac{2}{k}\psi^3 \psi^-~.
\ee
In the new basis, the non-vanishing operator product expansions of the fermions read:
\begin{align}
\psi^+(z) \psi^-(0) &\approx \frac{k}{z} \\
\psi^3(z)\psi^3(0) &\approx -\frac{k}{2z} ~. 
\end{align}
The operator product expansions between the currents and the fermions take the   form:
\begin{align}
\label{jpsioperator product expansion}
\hat j^+(z)\psi^+(0) \approx O(z)~,\quad 
\hat j^+(z)\psi^3(0)&\approx -\frac{\psi^+(0)}{z}~,\quad 
\hat j^+(z) \psi^-(0)\approx -\frac{2\psi^3(0)}{z}\cr
\hat j^3(z)\psi^+(0) \approx \frac{\psi^+(0)}{z}~,\quad 
\hat j^3(z)\psi^3(0) &\approx O(z)~,\quad\hspace{.7cm} 
\hat j^3(z)\psi^-(0)\approx -\frac{\psi^-(0)}{z}\cr
\hat j^-(z)\psi^+(0)\approx ~\frac{2\psi^3(0)}{z},\quad 
\hat j^-(z)\psi^3(0) &\approx \frac{\psi^-(0)}{z}~,\quad\hspace{.4cm} 
\hat j^-(z)\psi^-(0) \approx O(z)~.
\end{align}
The metric on the tangent space in the $(+,-,3)$ basis is off-diagonal:
\be
g_{AB} = \begin{pmatrix}
0 & \frac12 & 0 \\
\frac12 & 0 & 0 \\
0 & 0 & -1 
\end{pmatrix}~.
\ee
Thus we have $\psi_+ = \frac12 \psi^-$ and $\psi_- = \frac12 \psi^+$, $\psi_3 = -\psi^3$ and similarly for the currents. 
The stress tensor and its ${N}=1$ superpartner for the supersymmetric $AdS_3$ factor are 
\begin{align}
    T = \frac{1}{k}j^Aj_A - \frac{1}{k}\psi^A\psi_A ~,\qquad G = \frac{2}{k}(\psi^Aj_A + \frac{2\ii}{k} \psi^1\psi^2\psi^3)~.
\end{align}
These are standard structures in $N=1$ supersymmetric Wess-Zumino-Witten models.

\subsection{The  Circle Sector}

In addition to the three world sheet fermions associated to the space tangent to the three-dimensional anti-de Sitter space, we introduce a fermion $\psi^0$ that is tangent to the circle $S^1$. Its bosonic partner $J^0$  is the world sheet current that arises from translation invariances along the circle associated to the space-time U$(1)_R$ charge. These world sheet fields have the operator product expansions
\begin{align}
J^0(z) J^0(w) &\approx \frac{1}{(z-w)^2}\\
\psi^0(z) \psi^0(w) &\approx \frac{1}{z-w}\\	
J^0(z) \psi^0(w) &\approx 0 ~.
\end{align}  
This factor world sheet conformal field theory is free. 

\subsubsection{Spacetime R-charge from the Circle Direction}

We recall that the space-time  ${\mathcal N}=2$ superalgebra  includes a U$(1)_R$ current. The spacetime R-charge operator which we denote by ${\cal Q}_0$, corresponds to the zero mode along the $S^1$ direction \cite{Giveon:2003ku}
 \be
 \label{spacetimeRcharge}
 {\cal Q}_0 = \sqrt{2k} \oint J^0 = \oint e^{-\psi} \psi^0~.
 \ee
 When we bosonize the current $J^0(z)$  :
 \be
 J^0 = \ii \pa X~,
 \ee
 any vertex operator on the world sheet with spacetime R-charge $Q_{\text{st}}$ can  be represented in the form 
 \be
 V_{Q_{\text{st}}} = e^{\ii \frac{Q_{\text{st}}}{\sqrt{2k}}X } ~.
 \ee
 We make good use of this fact in the bulk of the paper.

\subsubsection{The Bosonization of Four Fermions} 
 
The four fermions can be paired up and it is useful to bosonize them in order to write down vertex operators in the Ramond sector. We define bosons $H_{0,1}$ through the formulas 
\be
\pa H_1 = - \frac{2}{k}\psi^1\psi^2~, \qquad \pa H_0 = -\ii \sqrt{\frac{2}{k}} \psi^0\psi^3~.
\ee
The bosons $H_I$ satisfy the  operator product expansions
\be
H_I(z) H_J(w) \approx -\delta_{IJ} \ln (z-w)\quad\text{for}\quad I,J = 0, 1~.
\ee
The fermions are exponentials in the bosons  :
\begin{align}
\frac{1}{\sqrt{k}}\psi^{\pm} = \frac{1}{\sqrt{k}} (\psi^1 \pm \ii \psi^2) &= e^{\pm \ii H_1} \\ 
\frac{1}{\sqrt{2}}\psi^0 \mp \frac{1}{\sqrt{k}}\psi^3 &= e^{\pm \ii \pi N_1} e^{\pm \ii H_0} ~,
 \end{align} 
 where we have included the cocycle factor that depends on the number operator  $N_1= \ii \oint \pa H_1$. This factor ensures that the fermion vertex operators anti-commute appropriately. Their presence invites us to  define hatted scalars 
 \be
 \label{hattedbosons}
 \hat H_1 = H_1 \qquad \hat H_0 = H_0 + \pi~ N_1~.
 \ee
In terms of the hatted scalars,  the bosonization formulae simplify. The fermionic currents $\hat j^A$ can be written as:
\begin{equation}
\hat j^3 = \ii\, \pa \hat H_1~, \qquad \hat j^{\pm} = \pm 
e^{\pm \ii \hat H_1}( e^{-\ii \hat H_0} - e^{+\ii \hat H_0})~.     
\end{equation}
Thus, we have conveniently expressed essential ingredients in the $AdS_3 \times S^1$ conformal field theories in terms of bosons only. 

\subsection{Operator Products in the Spin-field Basis}

In this section we list the operator product expansions between the spin fields and fermions used in the main text. We define a column matrix $S^\alpha$ which spans the vector space of the four-dimensional spin fields:
\be
S^\alpha \leftrightarrow
 \begin{pmatrix}
e^{\frac{\ii}{2} \hat H_1 + \frac{\ii}{2} \hat H_0} \\
e^{-\frac{\ii}{2} \hat H_1 + \frac{\ii}{2} \hat H_0} \\
e^{+\frac{\ii}{2} \hat H_1 - \frac{\ii}{2} \hat H_0} \\
e^{-\frac{\ii}{2} \hat H_1 - \frac{\ii}{2} \hat H_0}
\end{pmatrix} \, .
\ee 
The spin fields represent the Clifford algebra of the fermion zero modes:
\begin{align}
\label{psiSope}
\frac{1}{\sqrt{k}}\psi^{m}(z) S^\alpha(w) &\approx \frac{{(\gamma^m)^{\alpha}}_\beta S^\beta}{(z-w)^{\frac12}} \, .
\end{align}
The spin fields multiply to form the charge conjugation matrix $C$:
\begin{align}
\label{SSope}
S^{\alpha}(z)S^{\beta}(w) &\approx \frac{C^{\alpha \beta}}{(z-w)^{\frac12}} ~.
\end{align}
The $so(3,1)$ Clifford algebra 
\be
\{\gamma^m, \gamma^n\} = 2\eta^{mn}~,
\ee
with metric $\eta = \text{diag}(+,+,-,+)$ is explicitly represented by 
\begin{align}
\gamma_1= \sigma_1\otimes \mathbb{I}_2~,\quad
\gamma_2=\sigma_2\otimes \mathbb{I}_2~,\quad
\gamma_3=-\ii\sigma_3\otimes \sigma_2~,\quad
\gamma_4=\sigma_3\otimes \sigma_1~,
\end{align}
where $\sigma_a$ are the usual Pauli matrices and $\mathbb{I}_2$ is the two by two identity matrix.
The 
charge conjugation matrix ${C}$ satisfies
\begin{equation}
{C}\,\gamma_m\,{C}^{-1}= -(\gamma_m)^{T}
\end{equation}
where $T$ denotes the transpose. The charge conjugation matrix in our representation is:
\begin{equation}
{C}=\sigma_2\otimes \sigma_1 = 
\begin{pmatrix}
0 & 0 & 0 & -1\\
0 & 0 & -1 & 0\\
0 & 1& 0 & 0 \\
1 & 0 & 0 & 0
\end{pmatrix} \, .
\label{chargeconj4}
\end{equation}
To obtain these results one needs to take  care in choosing appropriate branch cuts for the spin fields and their operator product expansions. For example, the leading term in a spin field  operator product $e^{i \frac{\hat H}{2}}(z) e^{-i\frac{\hat H}{2}}(w) \approx (z-w)^{-\frac{1}{4}}e^{ \frac{i}{2} (\hat H(z)-\hat H(w))} $ picks up a fourth root of $-1$ when we exchange the positions of the operators. These roots must be chosen consistently in order to reproduce the claimed operator expansions for the spin fields.

\subsection{The Operators and Operator Products in the \texorpdfstring{$sl(2,\mathbb{R})$}{sl2r} Model }

An affine primary field $\Phi_h(x; z)$ of spin $h$ with respect to the  $sl(2,\mathbb{R})$ current algebra satisfies the expansion:
\be
\label{jPhiws}
j^A(z) \Phi_h(x; z) \approx 
-\frac{D_x^A \Phi_h(x; z)}{z-w}~,
\ee
where the differential operators $D^A_x$ are defined as
\be
\label{dbydx}
D^-_x = \pa_x~, \qquad D^3_x = x\frac{\pa}{\pa x} +h    ~,\qquad D^+_x = x^2\pa_x +2\, h\, x~. 
\ee
The differential operators $D^A_x$ represent the global $sl(2,\mathbb{R})$ subalgebra. Using the variable $x$ is a handy way to parameterize states in $sl(2,\mathbb{R})$ representations. For instance, 
it is possible to form $x$ dependent linear combinations of the currents, fermions and spin fields that render manifest their transformation properties under the    $sl(2,\mathbb{R})$ symmetry. One defines \cite{Kutasov:1999xu}
\be
j(x; z) = - j^+(z) + 2x j^3(z) - x^2j^+(z)~.
\ee
Using equation \eqref{jPhiws} one verifies that
\be
\label{phijreversed}
j(x_1;z) \Phi_h (x_2;w)  \approx  \frac{1}{z-w} ((x_1-x_2)^2 \partial_{x_2} - 2 h (x_1-x_2)) \Phi_h(x_2;w)~.
\ee
A similar combination of currents can also be used to define $\hat{j}(x; z)$ and $J(x; z)$ for the fermionic and supersymmetric currents respectively -- indeed, they transform in the same three-dimensional representation. The operator product expansion between the currents can be rewritten as 
\be
J(x_1, z) J(x_2, w) \approx k\frac{(x_1-x_2)^2}{(z-w)^2} + \frac{1}{z-w}((x_1-x_2)^2 \pa_{x_2}+ 2(x_1-x_2))J(x_2, w)~.
\ee
%
 The three fermions can similarly be combined: 
\be
\psi(x; z) = - \psi^+(z) + 2x\psi^3(z) - x^2\psi^-(z)~.
\ee
Then, first of all one can verify the operator product expansion 
\be
\label{jpsiws}
\hat{j}^A(z) \psi(x; z) \approx 
-\frac{D_x^A \psi(x; z)}{z-w}~,
\ee
where the $D_x^A$ are as in \eqref{dbydx} with $h=-1$. This shows that $\psi(x)$ is an $sl(2,\mathbb{R})$ primary field with spin  equal to $-1$. One can then rewrite the fermion operator product expansions:
\begin{align} 
 \psi(x_1;z_1) \psi(x_2;z_2) & \approx  k \frac{(x_{12})^2}{z_{12}} \\
\hat{j}(x_1; z) \psi(x_2, w) &\approx \frac{1}{z-w}((x_1-x_2)^2\pa_{x_2} + 2(x_1-x_2))\psi(x_2, w)~.
\end{align}
For the spin fields, we define the   linear combinations:  
\begin{align}
S(x; z) &= x\, e^{-\frac{\ii}{2} \hat H_1 + \frac{\ii}{2} \hat H_0} +  e^{\frac{\ii}{2} \hat H_1 - \frac{\ii}{2} \hat H_0}  \\
\widetilde S(x; z) &= - x\, e^{-\frac{\ii}{2} \hat H_1 - \frac{\ii}{2} \hat H_0} +  e^{\frac{\ii}{2} \hat H_1 + \frac{\ii}{2} \hat H_0}~. 
\end{align}
One can check that these transforms as spin $h=-\frac12$ fields. The fields $S$ and $\tilde{S}$ have opposite four-dimensional chirality. Using  the  operator product expansions \eqref{psiSope} and \eqref{SSope}, one derives the operator product expansions for the fields in the $x$-basis:
\begin{align}
    S(x_1; z)S(x_2; w) &\approx -\frac{x_1-x_2}{(z-w)^{\frac12}}\\
    \frac{1}{\sqrt{k}}\psi(x_1; z)S(x_2; w) &\approx 2(x_1-x_2) \frac{\widetilde S(x_2; w)}{(z-w)^{\frac12}} ~.
\end{align}

\subsubsection{The Operator Products of Bosonic Primary Fields}
\label{H3StructureConstants}

The primary fields $\Phi_h$  satisfy the mutual operator product expansion  -- to avoid clutter we are only presenting the holomorphic dependence on $x_1-x_2$ and $z-w$ :
 \begin{align}
 \label{phiphireversed}
 \Phi_{h_1} (x_1;z) \Phi_{h_2} (x_2;w) &\approx  
  \int d h \,  
\frac{(x_1-x_2)^{h-h_1-h_2}}{ (z-w)^{-\frac{h_1(h_1-1)}{k} -\frac{h_2(h_2-1)}{k}+\frac{h(h-1)}{k}}} {C_{h_1 h_2}}^{h} \, \Phi_{h} (x_2;w)~
+ \dots
\end{align}
The $AdS_3$ structure constant ${C_{h_1 h_2}}^{h_3}$ is given by
\begin{align}
\label{AdS3StructureConstant}
    {C_{h_1 h_2}}^{h_3} = \frac{\nu^{1-h_1-h_2-h_3}\Upsilon(b)}{\Upsilon(b(h_1+h_2+h_3)-b)} \frac{\Upsilon(2bh_1-b)\Upsilon(2bh_2-b) \Upsilon(2bh_3)}{\Upsilon(bh_1+bh_2-bh_3)\Upsilon(bh_2+bh_3-bh_1)\Upsilon(bh_3+bh_1-bh_2)}
\end{align}
where $b^2=\frac1k$, $\nu$ is a constant, and  $\gamma(x) = \frac{\Gamma(x)}{\Gamma(1-x)}$.  The special function $\Upsilon$ introduced in \cite{Zamolodchikov:1995aa} satisfies special shift properties 
\begin{equation}
    \Upsilon(x+b) = b^{1-2bx}\, \gamma(bx)\, \Upsilon(x)~,\quad \Upsilon(x+\frac{1}{b})= b^{-1+\frac{2x}{b}}\gamma(\frac{x}{b})\, \Upsilon(x)~,
\end{equation}
whenever $x$ lies outside the range $0< \text{Re}(x) < b+\frac{1}{b}$. The structure constant is obtained from the 
three point functions $C_H(h_1, h_2, h_3)$ of the theory on Euclidean $AdS_3$ derived in \cite{Teschner:1997ft} by the relation:
\be
{C_{h_1 h_2}}^{h_3} = \frac{C_H(h_1, h_2, h_3)}{C_H(1, h_3, h_3)}~.
\ee
We will be interested in the form of the operator product expansion coefficient for two particular cases: 
\paragraph{Case 1:} For $h_3=h_1+h_2-1$ we find the structure constant
\begin{equation}
{C_{h_1 h_2}}^{h_1+h_2-1} = 1 \, . \label{AStructureConstant}
\end{equation}
It is crucial that this is a universal constant independent of the spins of the primaries. 

\paragraph{Case 2:} For $h_3=h_1+h_2-2$ we find the structure constant:
 \begin{equation}
{C_{h_1 h_2}}^{h_1+h_2-2} = 
\frac{1}{\nu \gamma(\frac1k)}\frac{\gamma(\frac{(2h_1-2)}{k})\gamma(\frac{(2h_2-2)}{k})}{\gamma(\frac{2h_3}{k})} \, . \label{ASecondStructureConstant}
\end{equation}
This structure constant  depends on the spins.

\subsection{\texorpdfstring{The Operator Product Coefficients of $su(2)$}{Structure Constants of su(2)}}
The operator product coefficients of $su(2)$ at level $k$ can  be gleaned from \cite{Dabholkar:2007ey} after a convenient renormalization of the operators:
\begin{equation}
{{C^{su(2)}}_{j_1 j_2}}^{j_3} 
= {N_{j_1j_2}}^{j_3} P(j+1)
 \prod_{i=1}^3 \frac{P(j-2j_i)}{P(2j_i)} \frac{P(2j_3)}{P(2j_3+1)} \, .
\label{SU(2)StructureConstant}
\end{equation}
Here the function $P(s)$ is defined as a finite product:
\begin{align}
P(s) = \prod_{n=1}^s \gamma(\frac{n}{k})~,
\end{align}
with $P(0) = 1$, and $j=j_1+j_2+j_3$. As noted in \cite{Dabholkar:2007ey}, whenever $2h_i$ and $h_1+h_2+h_3$ are integer valued, the structure constants (\ref{AdS3StructureConstant}) and (\ref{SU(2)StructureConstant}) multiply into a constant when $j_i=h_i-1$. In the bulk of the paper, we encounter a further generalization of this property.
We also exploit the fact that the structure constants of the super parafermions are essentially given by those of the $su(2)$ Wess-Zumino-Witten model  \cite{Zamolodchikov:1986gh}.

 \section{ \texorpdfstring{Extended Superconformal Field Theory}{Y} }
 \label{TheoryY}
 \label{YTheory}

 The conformal field theory  $Y$ is an ${N}=(2,2)$ superconformal field theory with central charge $c_Y = 9-6/k$. The left moving superconformal symmetry is generated by the stress tensor $T$, the supercurrents $G^{\pm}$ and a U$(1)_R$ current $J_R$. In this Appendix we omit the explicit $Y$ indices 
 that would denote the fact that symbols refer to the internal conformal field theory $Y$ alone. To set the stage, we begin with a brief review of  (anti-)chiral rings and Ramond ground states. 

\subsection{Chiral Primaries, Spectral Flow and a Real Structure}
Chiral and anti-chiral primary states are defined by the conditions \cite{Lerche:1989uy}:
\begin{align}
G^{+}_{-\frac12} V_{c.p.}^a &= 0~, \qquad G^{\pm}_{n+\frac12} V_{c.p.}^a=0~, \quad\text{for}\quad n\ge 0\\
G^{-}_{-\frac12} V_{a.c.p.}^{\bar a} &= 0~, \qquad G^{\pm}_{n+\frac12} V_{a.c.p.}^{\bar a} =0~, \quad\text{for}\quad n\ge 0~.
\end{align}
From the supersymmetry algebra and unitarity one can check that they satisfy the extremal conditions $h^a = \frac{q^a}{2}$ and $h^{\bar a} = -\frac{q^{\bar a}}{2}$ respectively.  These operators have regular operator product expansions and one can define a chiral and anti-chiral ring  \cite{Lerche:1989uy}:
\begin{align}
      V^{ a}_{c.p.}(z)V^{ b}_{c.p.}(w) &\approx {R^{ a b}}_{ c} V^{ c}_{c.p.}(w)~,\\
     V^{\bar a}_{a.c.p.}(z)V^{\bar b}_{a.c.p.}(w) &\approx {R^{\bar a\bar b}}_{\bar c} V^{\bar c}_{a.c.p.}(w)~,
\end{align}
where the ring structure constants ${R^{ab}}_c$ and ${R^{\bar a\bar b}}_{\bar c}$ are related by complex conjugation (namely the assumed $\mathbb{Z}_2$ symmetry that charge conjugates chiral into anti-chiral primaries). 

In this section we write down the basic operator product expansions between chiral primaries, anti-chiral primaries and Ramond ground states in the $Y$-theory. There are a few important ingredients that enter the derivations. The first is the existence of spectral flow in  ${N}=2$ theories. Spectral flow by an amount $\eta$ acts on operators and shifts their R-charges and conformal dimensions according to the formulae:
\begin{equation}
J_{R,0} \rightarrow J_{R,0} + \frac{c}{3} \eta \, ,
\qquad
L_m \rightarrow L_m + \eta J_{R,m} + \frac{c}{6} \eta^2 \delta_{m,0} \, .
\end{equation}
The $J_{R,m}$ and $L_m$ are the  modes of the R-current and stress tensor respectively. The NS and Ramond sectors are related by spectral flow by an amount $\eta=\pm\frac12$  \cite{Lerche:1989uy}. In particular, Ramond ground states are related by spectral flow by  $\eta=\pm \frac12$ to (anti-)chiral primary operators in the NS sector.

The second ingredient is the bosonization of the R-current in terms of a canonically normalized scalar $Z$:
 \be
 J_R(z) = \sqrt{\frac{c}{3}}\,\ii\, \pa Z~.
 \ee
 Thus, an operator with worldsheet R-charge $q$ can be represented as the product
 \be
 O_{q} = e^{\ii \sqrt{\frac{3}{c}}q\, Z}\, \widetilde{O}~,
 \ee
 where the operator $\widetilde{O}$ has zero world sheet R-charge. 
  We shall use this fact extensively. An (anti-) chiral primary of charge $q^a$ ($q^{\bar a}$) can therefore be written in the product form:
\begin{equation}
\label{cpfactorized}
V_{c.p.}^a = e^{ i \sqrt{\frac{3}{c}} q^a Z} \Pi^a~,\qquad V^{\bar{a}}_{a.c.p.} = e^{ i \sqrt{\frac{3}{c}} 
q^{\bar{a}} Z} \Pi^{\bar{a}}  \, .
\end{equation}
We assume that the operators $\Pi^a, \Pi^{\bar a}$ decouple from the $U(1)_R$ sector. Note that the unique generalized volume operator of charge $\frac{c}{3}$ is purely in the $U(1)_R$ sector.

The third ingredient is a useful way to think of the connection between NS and Ramond sectors of $(2,2)$ superconformal theories that is advanced in \cite{Cecotti:1991me}. If we start with the lowest R-charge Ramond ground state, we can generate all other ground states by acting on it with the chiral ring:
\begin{align}
\Sigma^a &= V^a_{c.p.} \cdot \Sigma_0 \cr
&= e^{ \ii \sqrt{\frac{3}{c}} (q^a -\frac{c}{6}) Z}\, \Pi^a~,
\label{RRVcpreln}
\end{align}
where we have used the fact that the Ramond ground state $\Sigma_0 =e^{-\ii\sqrt{\frac{c}{12}}Y}$ is purely written in terms of the U$(1)_R$ sector. In the final expression in \eqref{RRVcpreln} we also see how the Ramond ground state is related to the chiral primary in the NS sector by spectral flow by half a unit. Alternatively if we start with the highest R-charge Ramond sector ground state, we generate all other ground states by acting with the anti-chiral ring:
\begin{align}
\Sigma^{\bar a} &= V^{\bar a}_{a.c.p.}  \cdot \bar \Sigma_0\cr
&= e^{\ii \sqrt{\frac{3}{c}} (q^{\bar a} +\frac{c}{6}) Z}\, \Pi^{\bar a}~,
\end{align}
where $\bar \Sigma_0 = e^{\ii\sqrt{\frac{c}{12}}Z}$. 
This means that Ramond ground states can be canonically associated to chiral ring elements ($a$) or to anti-chiral ring elements ($\bar{a}$). 

The last ingredient is a relation between these two choices of basis of Ramond ground states. They are related by a  real structure,  ${M^a}_{\bar a}$  \cite{Cecotti:1991me}: 
\be
\Sigma^{a} = {M^a}_{\bar a}~ \Sigma^{\bar a}~. 
\ee
The basis change ${M^a}_{\bar a}$ carries R-charge equal to $2q^a - \frac{c}{3}$. One can similarly define its complex conjugate  ${M^{\bar{a}}}_a$  which has R-charge $-2q_a + \frac{c}{3}$. The complex conjugate matrix equals the inverse matrix:
\begin{align}
{M^{a}}_{\bar a} {M^{\bar a}}_b  &= {\delta^a}_b~.
\end{align}
Given the real structure $M$ and the fact that the chiral and anti-chiral primaries are related to the Ramond ground states by half a unit of spectral flow, we also obtain another basis of (anti-)chiral primaries  labelled by $\bar a$ ($a$):
\begin{align}
V^{a}_{a.c.p.} &=  \text{Flow by minus one unit}(V^{a}_{c.p.})= {M^a}_{\bar{a}}  V^{\bar a}_{a.c.p.} \nonumber
\\
 V^{\bar{a}}_{c.p.} &=  \text{Flow by one unit}(V^{\bar{a}}_{a.c.p.})= {M^{\bar{a}}}_a  V^a_{c.p.} ~.
 \label{flowedcp}
\end{align}
These operators have charges $\mp (\frac{c}{3} - q_a)$ respectively. Given these ingredients, we are ready to derive the necessary operator product expansions starting from the chiral ring relations.

\subsection{Operator Products and a Metric}
Our derivation of various terms in operator product expansions is based on the chiral ring relation
\begin{align}
\label{cpringrelation}
V_{c.p.}^a (z) V_{c.p.}^b (w) & \approx  {R^{ab}}_c V_{c.p.}^c (w) + O(z-w)~.
\end{align}
Firstly, we derive an analogue of the chiral ring relation in which we factor out the $U(1)_R$ charge of the chiral primary operators. 
We  use the form \eqref{cpfactorized} of the chiral ring elements  and write the product of operators as:
\begin{align}
e^{ \ii \sqrt{\frac{3}{c}} q^a Z} \Pi^a (z)
e^{ \ii \sqrt{\frac{3}{c}} q^b Z} \Pi^b  (w)
& \approx (z-w)^{\frac{3}{c} q^a q^b} e^{\ii \sqrt{\frac{3}{c}} (q^a Z(z)+q^bZ(w)}
\Pi^a(z) \Pi^b(w) ~.
\end{align}
In order to be consistent with the ring relation  \eqref{cpringrelation} we  identify 
\begin{align}
\label{Vcdefn}
V_{c.p.}^c(w)  & = e^{i \sqrt{\frac{3}{c}} (q^a +q^b) Z(w))} \Pi^c(w)\, .
\end{align}
and reconstruct the leading term in the operator product expansion that the fields $\Pi^a$ must satisfy:
\begin{equation}
\Pi^a (z) \Pi^b(w) \approx (z-w)^{-\frac{3}{c} q^a q^b} ({R^{ab}}_c \Pi^c(w) + O(z-w)) \, .
\end{equation}
We can think of the fields $\Pi^a$ as R-parafermions. 
Now that we have the $\Pi^a$ operator product expansion, we more easily  obtain the leading term in the operator product expansion of the Ramond ground states:
\begin{align}
\Sigma^a(z) \Sigma^b(w) & \approx 
e^{ i \sqrt{\frac{3}{c}} (q^a-\frac{c}{6}) Z} \Pi^a  (z) e^{ \ii \sqrt{\frac{3}{c}} (q^b-\frac{c}{6}) Z} \Pi^b  (z) \nonumber \\
&\approx  (z-w)^{-\frac12 (q^a+q^b)+\frac{c}{12}}
\Big( e^{\ii \,  \sqrt{\frac{3}{c}} \big(q^a+q^b - \frac{c}{3}\big) Z}
{R^{ab}}_c \Pi^c(w) + O(z-w) \Big) 
\label{preRRacp} \, .
\end{align}
We have assumed that the charges $q^a$ corresponds to the spectrum of R-charges of the chiral primaries -- they  are therefore positive. We once more 
define the chiral primary operator $V^c_{c.p.}$ as in \eqref{Vcdefn} and compute:
\begin{align}
\Sigma^a(z) \Sigma^b(w)
&\approx  (z-w)^{-\frac12 (q^a+q^b)+\frac{c}{12}}
\Big( e^{-i \,  
 \sqrt{\frac{c}{3}} Z}~
{R^{ab}}_c~ V_{c.p.}^c(w)  + O(z-w) \Big) \nonumber \\
&\approx  (z-w)^{-\frac12 (q^a+q^b)+\frac{c}{12}}
\Big( 
{R^{ab}}_c \Big[ {M^c}_{\bar c}V_{a.c.p.}^{\bar c} \Big]+ O(z-w) \Big) 
\label{RRacp} \, .
\end{align}
In going from the first to the second line we observe that we have spectrally flowed  a chiral primary by minus one unit and this leads to the anti-chiral primary
(\ref{flowedcp}).  
We  note that the dimension of the anti-chiral primary with the non-canonical labelling $V^c_{a.c.p.}$ is given by $ \frac{c}{6}-\frac{ q^c}{2}  $.

A second equation can similarly be found by starting out with Ramond ground states labelled by anti-chiral primaries and performing the operator products:
\begin{align}
\Sigma^{\bar{a}}(z) \Sigma^{\bar{b}} (w) & \approx 
(z-w)^{\frac12(q^{\bar{a}}+q^{\bar{b}})+\frac{c}{12}}
\Big( 
{{R}^{\bar a\bar b}}_{\bar c}\left[ {M^{\bar c}}_{c} V_{c.p.}^{c} \right]+ O(z-w) \Big) \, .
\label{RRcp}
\end{align}
We find a chiral primary obtained by unit spectral flow from an anti-chiral primary, as  in equation \eqref{flowedcp}, and it has world sheet dimension $\frac{c}{6}+
\frac{q^{\bar{c}}}{2}$. 

We summarize that in the operator product of Ramond ground states one can get either a chiral primary or an anti-chiral primary as the leading term depending on the R-charges of the Ramond ground states involved. The relevant formulas  are  (\ref{RRacp}) and (\ref{RRcp}).

Since the Ramond ground state can be written as a (anti-)chiral ring element times the states with lowest (highest) R-charge, it is straightforward to compute the leading term in the operator product expansion between a (anti-)chiral primary and a Ramond ground state: 
\begin{align}
V^a_{c.p.} (z) \Sigma^b(w) & \approx e^{ i \sqrt{\frac{3}{c}} q^a Z} \Pi^a  (z) e^{ i \sqrt{\frac{3}{c}} (q^b-\frac{c}{6}) Z} \Pi^b  (z)
\nonumber \\
&\approx (z-w)^{-\frac{q^a}{2}} e^{ i \sqrt{\frac{3}{c}} (q^a+q^b-\frac{c}{6}) Z} {R^{ab}}_c \Pi^c(w) + \dots
\nonumber \\
& \approx \frac{ {R^{ab}}_c \Sigma^c(w)}{(z-w)^{\frac{q^a}{2}}} + \dots
\end{align}
Again, this holds whenever the structure constant is non-zero. Similarly we find that 
\begin{align}
V^{\bar{a}}_{a.c.p.} (z) \Sigma^{\bar{b}}(w) 
& \approx \frac{ {R^{\bar a\bar b}}_{\bar c} \Sigma^{\bar{c}}(w)}{(z-w)^{-\frac{q^{\bar{a}}}{2}}} + \dots
\end{align}
Lastly we derive the operator products between the chiral and anti-chiral ring elements. 
The two point function of a chiral and anti-chiral primary operator is defined to be
\begin{equation}
\langle V_{c.p.}^a (z) {V_{a.c.p.}^{\bar{b}}} (w) \rangle = \frac{g^{a \bar{b}}}{(z-w)^{2h_a} } \, ,
\end{equation}
where $g^{a\bar b}$ is the  Zamolodchikov metric and we have chosen a basis in which it is diagonal in the sense that $q^{\bar b} = -q^a$. 
In order to obtain the operator product coefficient of interest we shall compute a three-point function in two different ways \cite{deBoer:2008ss}. The three point function of interest is given by
\be
\langle V_{c.p.}^a V_{a.c.p.}^{\bar b} V_{a.c.p.}^{\bar{c}} \rangle = \frac{X^{a\bar b \bar c}}{(z_a-z_b)^{q^b}(z_c-z_a)^{q^c}}~. 
\ee
Here we have used the usual definition of the three point function and  the charge conservation equation 
$q^a = q^b+q^c$. 
In the first manner of calculation, we use the chiral ring relation and obtain
\begin{align}
\langle V_{c.p.}^a V_{a.c.p.}^{\bar b} V_{a.c.p.}^{\bar{c}} \rangle   &= {R^{\bar b\bar c}}_{\bar d}~\langle V_{c.p.}^{a}  V^{\bar d}_{a.c.p.}  \rangle
 \cr
&= \frac{{R^{\bar b\bar c}}_{\bar d}~g^{a \bar{d}}}{(z_a-z_c)^{2h^a}} ~.
\end{align}
Thus we have fixed the structure constant to be $X^{ab\bar c} ={R^{\bar b\bar c}}_{\bar d}~g^{a \bar{d}}$. 
Equivalently we can use the operator product expansion
\begin{align}
\label{CpAcpcp}
V_{c.p.}^a(z_a) V_{a.c.p.}^{\bar{b}}(z_b) \approx \frac{ {D^{a \bar{b}}}_{c}}{(z_a-z_b)^{q^b}}~ V_{c.p.}^{c}(z_a)
\end{align}
and obtain
\begin{align}
\langle V_{c.p.}^a V_{a.c.p.}^{\bar b} V_{a.c.p.}^{\bar{c}} \rangle  &= 
\frac{{ D^{a\bar b}}_c}{(z_a-z_b)^{q^b}}\frac{g^{c\bar c}}{(z_a-z_c)^{q^c}} ~.
\end{align}
We therefore find -- see also Appendix E of \cite{deBoer:2008ss} --:
\be
{D^{a\bar b}}_c = g^{a \bar d}~ {R^{\bar b \bar e}}_{\bar d} ~ g_{c \bar e}~.
\ee
We can write this equality as $D=g R g^{-1}$ where we indicate with $g$ the Zamolodchikov metric with upper indices.

\section{The Spacetime Global Superconformal Algebra}
\label{SpacetimeSusyAlgebra}

The global ${\cal N}=2$ superconformal algebra  is described by the commutation relations:
\begin{align}
\label{spacetimesusyalgebra} 
\{ {\cal G}_r^+, {\cal G}_s^- \} &= 2{\cal L}_{r+s} + (r-s){\cal Q}_0 \cr
[{\cal L}_m, {\cal L}_n] &= (m-n) {\cal L}_{m+n}\cr
[{\cal L}_m, {\cal G}_r^{\pm} ] &= (\frac{m}{2}-r){\cal G}_{m+r}^{\pm}\cr
[{\cal Q}_0, {\cal G}_r^{\pm}] &=  \pm {\cal G}_r^{\pm}~,
\end{align}
for $r,s  = \pm \frac12$ and $ m,n = 0, \pm1$.
Using the currents of the worldsheet theory it is possible to write down integrated operators on the worldsheet that correspond to these global generators \cite{Giveon:2003ku}. The zero modes of the  bosonic currents generate the isometries of the target space. If we define
\begin{align}
{\cal L}_0 &= -\oint J^3 = -\oint e^{-\phi} \psi^3~,  \\ 
{\cal L}_{\pm} &= -\oint (J_1\pm  \ii J_2) = \oint e^{-\phi}(\psi_1 \pm \ii \psi_2)~,
\end{align}
then these generators satisfy the global  Virasoro algebra. The U$(1)_R$ charge in spacetime was defined in equation \eqref{spacetimeRcharge} 
 \be
 {\cal Q}_0 = \sqrt{2k} \oint J^0 = \oint e^{-\psi} \psi^0~.
 \ee
It therefore remains to write down the spacetime  supercharges. In the  $(-\frac12)$ picture they read: 
\be
\label{spacetimeQvop}
{\cal G}^{\pm}_r = \oint e^{-\frac12 \phi}~ S_r^{\pm}~,\quad \text{with}\quad r=  \pm\frac12~.
\ee
The operator $\phi$ is the bosonized superghost and the exponential has worldsheet conformal dimension $\frac{3}{8}$. Thus the vertex operator $S_r^{\pm}$ must have dimension equal to $\frac{5}{8}$. The key to writing down the vertex operators for the spacetime supercharges is to keep track of the charges under the bosonic spacetime symmetries. In particular from the spacetime algebra it is clear that the spacetime supercharges have $\pm 1$ charge under the spacetime $R$-current ${\cal Q}_0$ and must therefore include the factor $e^{\pm \frac{\ii}{\sqrt{2k}} X}$ in their vertex operators. Similarly, their Lorentz transformation properties fix their form to be
\be
S_r^{\pm} = e^{-\ii r (H_1\mp H_0)} e^{\pm \frac{\ii}{\sqrt{2k}} X}  \Sigma^{\pm}~,\quad \text{with}\quad r=  \pm\frac12~.
\ee
The operator factors $\Sigma^{\pm}$ are  purely in the $Y$-theory. By using the constraint that the worldsheet dimension of $S_r^{\pm}$ should be $\frac58$ we see that  
\be
\Delta(\Sigma^{\pm}) =   \frac{3}{8} - \frac{1}{4k}=\frac{c_Y}{24}~.
\ee
Thus we conclude that these must be Ramond ground states of the theory  $Y$. A priori there are many Ramond ground states in $Y$ but, as shown in \cite{Giveon:2003ku}, the Ramond ground states that appear in the spacetime supercharge are those with the highest or lowest  worldsheet R-charge $\pm \frac{c_Y}{6}$. These can be expressed purely in terms of the boson $Z$ that captures the world sheet $U(1)_R$ current direction: 
\be
S_r^{\pm} = e^{-\ii r (H_1\mp H_0)} e^{\pm \frac{\ii}{\sqrt{2k}} X}  e^{\pm \ii \sqrt{\frac{c_Y}{12}} Z} ~.
\ee
Using the free field operator products, one checks that the integrated vertex operators ${\cal L}_n, {\cal G}_r^{\pm}$ and ${\cal Q}_0$ indeed satisfy the spacetime ${\mathcal N}=2$ global supersymmetry algebra  \eqref{spacetimesusyalgebra}.

\bibliographystyle{JHEP}

\begin{thebibliography}{99}
\bibitem{tHooft:1993dmi}
G.~'t Hooft,
``Dimensional reduction in quantum gravity,''
Conf. Proc. C \textbf{930308} (1993), 284-296
[arXiv:gr-qc/9310026 [gr-qc]].
\bibitem{Susskind:1994vu}
L.~Susskind,
``The World as a hologram,''
J. Math. Phys. \textbf{36} (1995), 6377-6396
doi:10.1063/1.531249
[arXiv:hep-th/9409089 [hep-th]].
\bibitem{Maldacena:1997re}
J.~M.~Maldacena,
``The Large N limit of superconformal field theories and supergravity,''
Adv. Theor. Math. Phys. \textbf{2} (1998), 231-252
doi:10.1023/A:1026654312961
[arXiv:hep-th/9711200 [hep-th]].

\bibitem{Kutasov:1999xu}
D.~Kutasov and N.~Seiberg,
``More comments on string theory on AdS(3),''
JHEP \textbf{04}, 008 (1999)
doi:10.1088/1126-6708/1999/04/008
[arXiv:hep-th/9903219 [hep-th]].


\bibitem{Maldacena:2001km}
J.~M.~Maldacena and H.~Ooguri,
``Strings in AdS(3) and the SL(2,R) WZW model. Part 3. Correlation functions,''
Phys. Rev. D \textbf{65} (2002), 106006
doi:10.1103/PhysRevD.65.106006
[arXiv:hep-th/0111180 [hep-th]].
\bibitem{Dei:2021xgh}
A.~Dei and L.~Eberhardt,
``String correlators on $\text{AdS}_3$: Three-point functions,''
[arXiv:2105.12130 [hep-th]].




\bibitem{Li:2020nei}
S.~Li and J.~Troost,
``Twisted String Theory in Anti-de Sitter Space,''
JHEP \textbf{11} (2020), 047
doi:10.1007/JHEP11(2020)047
[arXiv:2005.13817 [hep-th]].


\bibitem{Giveon:2003ku}
A.~Giveon and A.~Pakman,
``More on superstrings in AdS(3) x N,''
JHEP \textbf{03} (2003), 056
doi:10.1088/1126-6708/2003/03/056
[arXiv:hep-th/0302217 [hep-th]].

\bibitem{Dabholkar:2007ey}
A.~Dabholkar and A.~Pakman,
``Exact chiral ring of AdS(3) / CFT(2),''
Adv. Theor. Math. Phys. \textbf{13} (2009) no.2, 409-462
doi:10.4310/ATMP.2009.v13.n2.a2
[arXiv:hep-th/0703022 [hep-th]].

\bibitem{Gaberdiel:2007vu}
M.~R.~Gaberdiel and I.~Kirsch,
``World sheet correlators in AdS(3)/CFT(2),''
JHEP \textbf{04} (2007), 050
doi:10.1088/1126-6708/2007/04/050
[arXiv:hep-th/0703001 [hep-th]].


\bibitem{Giveon:1999zm}
A.~Giveon, D.~Kutasov and O.~Pelc,
``Holography for noncritical superstrings,''
JHEP \textbf{10} (1999), 035
doi:10.1088/1126-6708/1999/10/035
[arXiv:hep-th/9907178 [hep-th]].

\bibitem{Argurio:2000tb}
R.~Argurio, A.~Giveon and A.~Shomer,
``Superstrings on AdS(3) and symmetric products,''
JHEP \textbf{12} (2000), 003
doi:10.1088/1126-6708/2000/12/003
[arXiv:hep-th/0009242 [hep-th]].

\bibitem{Eberhardt:2017fsi}
L.~Eberhardt, M.~R.~Gaberdiel, R.~Gopakumar and W.~Li,
``BPS spectrum on AdS$_3\times $S$^3 \times $S$^3 \times $S$^1$,''
JHEP \textbf{03} (2017), 124
doi:10.1007/JHEP03(2017)124
[arXiv:1701.03552 [hep-th]].

\bibitem{Ashok:2020dnc}
S.~K.~Ashok and J.~Troost,
``Superstrings in Thermal Anti-de Sitter Space,''
JHEP \textbf{04} (2021), 007
doi:10.1007/JHEP04(2021)007
[arXiv:2012.08404 [hep-th]].


\bibitem{Lerche:1989uy}
W.~Lerche, C.~Vafa and N.~P.~Warner,
``Chiral Rings in N=2 Superconformal Theories,''
Nucl. Phys. B \textbf{324}, 427-474 (1989)
doi:10.1016/0550-3213(89)90474-4


\bibitem{Friedan:1985ge}
D.~Friedan, E.~J.~Martinec and S.~H.~Shenker,
``Conformal Invariance, Supersymmetry and String Theory,''
Nucl. Phys. B \textbf{271}, 93-165 (1986)
doi:10.1016/0550-3213(86)90356-1

\bibitem{Polchinski:1998rr}
J.~Polchinski,
``String theory. Vol. 2: Superstring theory and beyond,'' Cambridge University Press 2001, 
doi:10.1017/CBO9780511618123
\bibitem{Maldacena:2000hw}
J.~M.~Maldacena and H.~Ooguri,
``Strings in AdS(3) and SL(2,R) WZW model 1.: The Spectrum,''
J. Math. Phys. \textbf{42} (2001), 2929-2960
doi:10.1063/1.1377273
[arXiv:hep-th/0001053 [hep-th]].


\bibitem{Giribet:2007wp}
G.~Giribet, A.~Pakman and L.~Rastelli,
``Spectral Flow in AdS(3)/CFT(2),''
JHEP \textbf{06}, 013 (2008)
doi:10.1088/1126-6708/2008/06/013
[arXiv:0712.3046 [hep-th]].



\bibitem{Cecotti:1991me}
S.~Cecotti and C.~Vafa,
``Topological antitopological fusion,''
Nucl. Phys. B \textbf{367}, 359-461 (1991)
doi:10.1016/0550-3213(91)90021-O



\bibitem{Aharony:2007rq}
O.~Aharony and Z.~Komargodski,
``The Space-time operator product expansion in string theory duals of field theories,''
JHEP \textbf{01} (2008), 064
doi:10.1088/1126-6708/2008/01/064
[arXiv:0711.1174 [hep-th]].

\bibitem{DiFrancesco:1997nk}
P.~Di Francesco, P.~Mathieu and D.~Senechal,
``Conformal Field Theory,''  Springer-Verlag New York (1997),
doi:10.1007/978-1-4612-2256-9.



\bibitem{Teschner:1999ug}
J.~Teschner,
``Operator product expansion and factorization in the H+(3) WZNW model,''
Nucl. Phys. B \textbf{571} (2000), 555-582
doi:10.1016/S0550-3213(99)00785-3
[arXiv:hep-th/9906215 [hep-th]].


\bibitem{deBoer:2008ss}
J.~de Boer, J.~Manschot, K.~Papadodimas and E.~Verlinde,
``The Chiral ring of AdS(3)/CFT(2) and the attractor mechanism,''
JHEP \textbf{03}, 030 (2009)
doi:10.1088/1126-6708/2009/03/030
[arXiv:0809.0507 [hep-th]].

\bibitem{Maldacena:2001ky}
J.~M.~Maldacena, G.~W.~Moore and N.~Seiberg,
``Geometrical interpretation of D-branes in gauged WZW models,''
JHEP \textbf{07} (2001), 046
doi:10.1088/1126-6708/2001/07/046
[arXiv:hep-th/0105038 [hep-th]].

\bibitem{Lunin:2000yv}
O.~Lunin and S.~D.~Mathur,
``Correlation functions for M**N / S(N) orbifolds,''
Commun. Math. Phys. \textbf{219} (2001), 399-442
doi:10.1007/s002200100431
[arXiv:hep-th/0006196 [hep-th]].

\bibitem{Zamolodchikov:1986gh}
A.~B.~Zamolodchikov and V.~A.~Fateev,
``Disorder Fields in Two-Dimensional Conformal Quantum Field Theory and N=2 Extended Supersymmetry,''
Sov. Phys. JETP \textbf{63} (1986), 913-919

\bibitem{Zamolodchikov:1995aa}
A.~B.~Zamolodchikov and A.~B.~Zamolodchikov,
``Structure constants and conformal bootstrap in Liouville field theory,''
Nucl. Phys. B \textbf{477}, 577-605 (1996)
doi:10.1016/0550-3213(96)00351-3
[arXiv:hep-th/9506136 [hep-th]].

\bibitem{Teschner:1997ft}
J.~Teschner,
``On structure constants and fusion rules in the SL(2,C) / SU(2) WZNW model,''
Nucl. Phys. B \textbf{546}, 390-422 (1999)
doi:10.1016/S0550-3213(99)00072-3
[arXiv:hep-th/9712256 [hep-th]].

\end{thebibliography}

\end{document}